\documentclass[aps,prl,twocolumn,preprintnumbers,10pt,showpacs]{revtex4-1}

\usepackage{amsmath,bm,amssymb}

\usepackage{graphicx}
\usepackage{psfrag}
\usepackage[caption=false]{subfig}

\usepackage{color,xcolor}
\definecolor{urlblue}{rgb}{0.2,0.4,0.7}
\definecolor{citegreen}{rgb}{0,0.6,0.2}
\definecolor{linkred}{rgb}{0.9,0.2,0.1}
\usepackage{hyperref}
\hypersetup{
colorlinks=true, citecolor=citegreen, linkcolor=blue,urlcolor=urlblue}

\usepackage{slashed}

\usepackage{tikz}
\usetikzlibrary{positioning,arrows}
\usetikzlibrary{decorations.pathmorphing}
\usetikzlibrary{decorations.markings}
\usetikzlibrary{shapes.geometric}
\tikzset{
    vector/.style={decorate, decoration={snake}, draw},
    provector/.style={decorate, decoration={snake,amplitude=2.5pt}, draw},
    antivector/.style={decorate, decoration={snake,amplitude=-2.5pt}, draw},
    fermion/.style={draw=black,
      postaction={decorate},decoration={markings,mark=at position .55
        with {\arrow[draw=black]{>}}}}, 
    fermionbar/.style={draw=black, postaction={decorate},
                       decoration={markings,mark=at position .55 with {\arrow[draw=black]{<}}}},
    fermionnoarrow/.style={draw=black},
    gluon/.style={decorate, draw=black,decoration={coil,amplitude=4pt, segment length=4pt}},
    scalar/.style={dashed,draw=black,
      postaction={decorate},decoration={markings,mark=at position .55
        with {\arrow[draw=black]{>}}}}, 
    scalarbar/.style={dashed,draw=black,
      postaction={decorate},decoration={markings,mark=at position .55
        with {\arrow[draw=black]{<}}}}, 
    scalarnoarrow/.style={dashed,draw=black},
    electron/.style={draw=black,
      postaction={decorate},decoration={markings,mark=at position .55
        with {\arrow[draw=black]{>}}}}, 
    bigvector/.style={decorate, decoration={snake,amplitude=4pt}, draw},
}

\allowdisplaybreaks

\begin{document}

\def\l{\left}
\def\r{\right}
\def\ep{\epsilon}
\def\bt{\beta}

\preprint{MITP/16-107}

\title{Konishi Form Factor at Three Loop in ${\cal N}=4$ SYM}

\author{Taushif Ahmed$^{a,b}$, Pulak Banerjee$^{a,b}$, Prasanna K.\
  Dhani$^{a,b}$, Narayan Rana$^{a,b}$, V. Ravindran$^{a}$ and Satyajit
  Seth$^{c}$} 

\affiliation{$^a$ The Institute of Mathematical Sciences, Taramani,
  Chennai 600113, India \\ $^{b}$ Homi Bhaba National Institute,
  Training School Complex, Anushakti Nagar, Mumbai 400085, India \\
$^c$ 
PRISMA Cluster of Excellence, Institut f\"{u}r Physik, Johannes
Gutenberg-Universit\"{a}t Mainz, D\,-\,55099 Mainz, Germany} 


\begin{abstract}
We present the first results on the third order corrections to
on-shell form factor (FF) of the Konishi operator     
in ${\cal N}=4$ supersymmetric Yang-Mills theory using Feynman diagrammatic approach in modified dimensional
reduction ($\overline {DR}$) scheme.  We show that it satisfies KG
equation in $\overline {DR}$ scheme while the result obtained in four
dimensional helicity (FDH) scheme needs to be suitably modified not
only to satisfy the 
KG equation but also to get the correct ultraviolet (UV) anomalous dimensions.   
We find that the cusp, soft and collinear anomalous dimensions obtained
to third order are same as those of  
the FF of the half-BPS operator confirming the universality of the infrared (IR) structures
of on-shell form factors.   
In addition, the highest transcendental terms of the FF of Konishi operator are identical to those of half-BPS operator 
indicating the probable existence of deeper structure of the on-shell FF.  
We also confirm the UV anomalous dimensions of Konishi operator up to
third order providing a consistency  
check on the both UV and universal IR structures in ${\cal N}=4$.

\end{abstract}

\maketitle

The ability to accomplish the challenging job of calculating
quantities is of fundamental importance in any potential mathematical
theory. In quantum field theory (QFT), this manifests itself in the quest for
computing the multi-loop and multi-leg scattering amplitudes under the
glorious framework of age-old perturbation theory. The fundamental
quantities to be calculated in any gauge theory are the scattering amplitudes or
the correlation functions. 
Recently, there have been surge of interest to study form factors (FFs) as they connect
fully on-shell amplitudes and correlation functions.
%In this letter, we confine ourselves in the
The FFs are a set of quantities which are constructed out of the
scattering amplitudes
involving on-shell states consisting of elementary fields and an
off-shell state described through a composite operator. These are
operator matrix elements of the form
$\langle p_1^{\sigma_1},\cdots,p_l^{\sigma_l}|{\cal O}|0 \rangle$
where, ${\cal O}$ represents a gauge invariant composite operator
which generates a multi-particle on-shell state
$|p_1^{\sigma_1},\cdots,p_l^{\sigma_l}\rangle$ upon operating on the
vacuum of the theory. $p_{i}$ are the momenta and $\sigma_i$
encapsulate all the other quantum numbers of the 
particles. More precisely, FFs are the amplitudes of the
processes where classical current or field, coupled through gauge
invariant composite operator ${\cal O}$, produces some quantum
state. Studying these quantities not only help to understand the
underlying ultraviolet and infrared structures of the theory, but also
enable us to calculate the anomalous dimensions of the associated
composite operator.    

The Sudakov FFs ($l=2$) in ${\cal N}=4$ maximally supersymmetric Yang-Mills (SYM) 
theory ~\cite{Brink:1976bc, Gliozzi:1976qd} were initially considered by van Neerven in~\cite{vanNeerven:1985ja}, almost
three decades back, where a half-BPS operator belonging to the stress-energy
supermultiplet, that contains the conserved currents of ${\cal N}=4$
SYM, was investigated to 2-loop order:
\begin{align}
\label{eq:1}
&{\cal O}_{\rm BPS} = \phi^{a}_{m} \phi^a_n - \frac{1}{3} \delta_{mn}
  \phi^a_s \phi^a_{s}\,.
\end{align}
 Very recently, this was
extended to 3-loop in~\cite{Gehrmann:2011xn}. We
will represent scalar and pseudo-scalar fields by $\phi^a_m$ and
$\chi^a_m$, respectively. The symbol $a \in [1,N^2-1]$ denotes the SU(N)
adjoint color index, whereas $m,n$ stand for the generation indices
which run from $[1,n_{g}]$. In $d=4$ dimensions, we have $n_g=3$. The sum over
repeated index will be assumed throughout the letter unless otherwise
stated. One of the most 
salient features of this operator is that, it is
protected by the supersymmetry (SUSY) i.e. the FFs
exhibit no ultraviolet (UV) divergences but infrared (IR) ones to all
orders in perturbation theory. In this article, our goal is to
investigate the Sudakov FFs of another very sacred operator in 
the context of ${\cal N}=4$ SYM, called Konishi operator, which is not
protected by the SUSY and consequently, exhibits UV divergences beyond
leading order:
\begin{align}
\label{eq:2}
&{\cal O}_{\cal K} = \phi^a_m \phi^a_m + \chi^a_m \chi^a_m\,.
\end{align}
The existence of UV divergences is captured through the presence of
non-zero anomalous dimensions. This operator is one of the members of
the Konishi supermultiplet and all the members of
the multiplet give rise to same anomalous dimensions. The one and two
loop Sudakov FFs of Konishi operator were computed
in~\cite{Nandan:2014oga} employing the on-shell unitarity method. In
addition, the IR poles at 3-loop were also 
predicted in the same article using the universal behaviour of those,
though the finite part was not 
computed. In this letter, we calculate the full 3-loop Sudakov
FFs using the age-old Feynman diagrammatic approach. In the
same spirit of the FFs in quantum chromodynamics (QCD), we examine the
results in the context of KG equation~\cite{Sudakov:1954sw,
  Mueller:1979ih, Collins:1980ih, Sen:1981sd}. Quite remarkably, it
has 
been found that the logarithms of the FFs satisfy the universal
decomposition in terms of the cusp, collinear, soft and UV anomalous
dimensions, exactly similar to those of QCD~\cite{Ravindran:2004mb,
  Moch:2005tm}! Except UV, which is a property of the associated
operator, all the remaining universal anomalous dimensions match
exactly with the leading transcendental terms of the corresponding ones in 
QCD upon putting $C_F=n_f T_{f}=C_A$. The quantities $C_{F}$ and
$C_{A}$ are the quadratic Casimirs of the SU(N) gauge group in
fundamental and adjoint representations, respectively. $n_f$ is the
number of active quark flavors and $T_{f}=1/2$.

\section{Framework of the Calculation}
\label{sec:framework}

The interacting Lagrangian encapsulating the interaction between
off-shell state ($J$) described by ${\cal O}_{\rm BPS}$ or ${\cal O}_{\cal
  K}$ and the fields in ${\cal N}=4$ SYM are given by
\begin{align}
\label{eq:3}
&{\cal L}^{\rm BPS}_{int} = J_{\rm BPS}^{mn} {\cal O}_{\rm BPS}\,, \quad 
{\cal L}^{\cal K}_{int} = J_{\cal K} {\cal O}_{\cal K}\,.
\end{align}
We define the form factors at ${\cal O}(a^{n})$ as
\begin{align}
\label{eq:4}
{\cal F}^{\rho,(n)}_{f} \equiv \frac{\langle {\cal M}^{\rho,(0)}_{f} |
  {\cal M}^{\rho,(n)}_{f} \rangle}{\langle {\cal M}^{\rho,(0)}_{f}|{\cal
  M}^{\rho,(0)}_{f}\rangle} 
\end{align}
where, $n=0,1,2,\cdots$ and $a$ is the `t Hooft
coupling~\cite{Bern:2005iz}:
\begin{align}
\label{eq:8}
a \equiv \frac{g_{\rm YM}^2 C_{A}}{\left( 4\pi \right)^2} \left( 4\pi
  e^{-\gamma_{E}} \right)^{-\frac{\epsilon}{2}}\,,
\end{align}
that depends on the Yang-Mills coupling constant $g_{\rm YM}$,
the loop-counting parameter and $C_{A}$.
The quantity $|{\cal M}^{\rho,(n)}_{f} 
\rangle$ is the transition matrix 
element of ${\cal O}(a^n)$ for the production of a pair of
on-shell particles $f{\bar f}$ 
from the off-shell state represented through $\rho$. For the case
under consideration, we take $f=\phi^a_m={\bar f}$, $\rho={\cal K}$ and
 ${\rm BPS}$ for $J_{\cal K}$ and $J^{mn}_{\rm BPS}$,
 respectively. The full form factor in terms of the
 components~(\ref{eq:4}) reads as
\begin{align}
\label{eq:5}
{\cal F}^{\rho}_{f} = 1+\sum\limits_{n=1}^{\infty} \left[ a^n \left(
  \frac{Q^2}{\mu^2} \right)^{n \frac{\epsilon}{2}} {\cal F}^{\rho,(n)}_{f} \right]\,.
\end{align}
The transition matrix element also follows same expansion. The
quantity $Q^{2}=-2 p_1.p_2$ and $\mu$ is introduced to keep the
coupling constant $a$ dimensionless in $d=4+\epsilon$ dimensions. 
%Our central \textbf{\textit{goal}} is to compute the ${\cal F}^{{\cal K},
%  (3)}_{\phi}$. 

\section{Regularization Prescriptions}
\label{sec:PresOfReg}

The calculation of the FFs in ${\cal N}=4$
SYM theory involves a subtlety originating from the dependence of the
composite operators on space-time dimensions $d$. Unlike
the half-BPS operator ${\cal O}_{\rm BPS}$, the Konishi
operator ${\cal O}_{\cal K}$ involves a sum over generation of the scalar and
pseudo-scalar fields and consequently, it does depend on $d$. The problem arises while
making the choice of regularization scheme~\cite{Nandan:2014oga}, which is necessary in
order to regulate the theory for identifying the
nature of divergences present in the FFs. Though the FFs of the
protected operators are free from UV divergences in 4-dimensions,
these do involve IR divergences arising from the soft and collinear
configurations of the loop momenta. 

For performing the regularization,
there exists several schemes, the four dimensional helicity
(FDH)~\cite{Bern:1991aq, Bern:2002zk} formalism is the most popular
one where everything is treated in 4-dimensions, except the loop
integrals that are evaluated in $d$-dimensions. In spite of its
spectacular applicability, this prescription may fail to produce the
correct result for the operators involving space-time dimensions~\cite{Nandan:2014oga}, such
as Konishi! However, this can be rectified and the rectification
scenarios differ from one operator to another. According to the
prescription prescribed in the article~\cite{Nandan:2014oga}, in order
to obtain the correct result for the
Konishi operator, one requires to multiply a factor of $\Delta^{\rm
  BPS}_{\cal K}$ which is  $\Delta_{\cal K}^{\rm BPS}= {n_{g,\epsilon}}/{2 n_{g}}$
%given by  
%
%\begin{align}
%\label{eq:6}
%\Delta^{\rm BPS}_{\cal K} = \frac{n_{g,\epsilon}}{n_{g}}
%\end{align}
%
with the difference between the FFs of the Konishi and those of
BPS i.e.
\begin{align}
\label{eq:7}
{\cal F}^{\cal K}_{f} = {\cal F}^{\rm BPS}_{f} + \delta {\cal F}_f^{\cal K}, 
\end{align}
where, 
$\delta {\cal F}_f^{\cal K}=\Delta^{\rm
  BPS}_{\cal K} ({\cal F}^{\cal K}_{f} - {\cal F}^{\rm BPS}_{f})$.
%\end{align}
%
The second subscript of $n_{g,\epsilon}$ represents the
dependence of the number of generations of the scalar and
pseudo-scalar fields on
the space-time dimensions: $n_{g,\epsilon}=(2 n_g-\epsilon)$. 
The prescription is validated through the production of the
correct anomalous dimensions up to 2-loop. In this article, for the
first time, this formalism is applied to the case of 3-loop FFs and is
observed to produce the correct anomalous dimensions for the Konishi. 

On the other hand, there exists another very elegant
formalism, called modified dimensional reduction ($\overline{\rm
  DR}$)~\cite{Siegel:1979wq, Capper:1979ns}, 
which is very much similar to the `t Hooft and Veltman prescription of
the dimensional regularization and quite remarkably, it is universally applicable to all
kinds of operators including the ones dependent on the space-time
dimensions. In this prescription, in addition to 
treating everything in $d=4+\epsilon$ dimensions, the number of
generations of the scalar and pseudo-scalar fields is considered as
$n_{g,\epsilon}/2$ instead of $n_{g}$ in order to preserve the ${\cal N}=4$ SUSY throughout. 
The dependence on $\epsilon$ preserves SUSY in a sense that the total number of
gauge, 
%($4+\epsilon$)
%, 
scalar ($n_{g}$) and pseudo-scalar ($n_{g}$)
degrees of freedom continues to remain 8. Within this framework, we
have calculated the Konishi FFs up to 3-loop level and the results
come out to be exactly same as the ones obtained in
Eq.~(\ref{eq:7}). This, in turn, provides a direct check on the
earlier prescription. In the next section, we will
discuss the methodology of computing the FFs.

\section{Calculation of the Form Factors}
\label{sec:CalcOfFF}

The calculation of the FFs follows closely the steps used
in the derivation of the 3-loop spin-2 and pseudo-scalar 
FFs in QCD~\cite{Ahmed:2015qia, Ahmed:2015qpa}.
In contrast to the most popular method of on-shell
unitarity for computing the 
scattering amplitudes in ${\cal N}=4$ SYM, we use the conventional
Feynman diagrammatic approach, which carries its own advantages in
light of following the regularization scheme, to 
accomplish the job. The relevant Feynman diagrams are generated
using QGRAF~\cite{Nogueira:1991ex}. Indeed, very special care is
taken to incorporate the Majorana fermions present in the ${\cal
  N}=4$ SYM appropriately. For Konishi as well as half-BPS operator,
1631 number of Feynman diagrams appear at 3-loop order which include the
scalar, pseudo-scalar, gauge boson as well as Majorana fermions in the
loops. The ghost loops are also taken into account ensuring the
inclusion of only the physical degrees of freedom of the gauge bosons. The raw
output of the QGRAF is converted to a suitable format for further calculation.
Employing a set of in-house routines based on Python and the symbolic
manipulating program FORM~\cite{Vermaseren:2000nd}, the simplification
of the matrix elements involving the Lorentz, color, Dirac and generation
indices is performed. In the FDH regularization scheme, except the
loop integrals all the remaining algebra is performed in $d=4$,
whereas in ${\overline{\rm DR}}$, everything is executed in
$d=4+\epsilon$ dimensions. While calculating within the framework of
${\overline{\rm DR}}$, the factor of $1/3$ in the second part of ${\cal
  O}_{\rm BPS}$, Eq.~(\ref{eq:1}), should be replaced by
$2/n_{g,\epsilon}$ to maintain its traceless property in $d$-dimensions.

The expressions involve thousands of apparently different 3-loop
Feynman scalar integrals. However, they are expressible in terms of a
much smaller set, called master integrals (MIs),
by employing the integration-by-parts (IBP)~\cite{Tkachov:1981wb,
  Chetyrkin:1981qh} and Lorentz invariance (LI)~\cite{Gehrmann:1999as}
identities. Though, the LI are not linearly independent of the
IBP~\cite{Lee:2008tj}, their inclusion however accelerates the
procedure of obtaining the solutions. All the scalar integrals are
reduced to the set of MIs using a Mathematica based package
LiteRed~\cite{Lee:2012cn, Lee:2013mka}. In the literature, there exists similar packages
to perform the reduction: AIR~\cite{Anastasiou:2004vj}, FIRE~\cite{Smirnov:2008iw},
Reduze2~\cite{vonManteuffel:2012np, Studerus:2009ye}. As a result, all
the thousands of scalar integrals can be expressed in terms of 22
topologically different MIs which were computed
analytically as Laurent series in $\epsilon$
in the articles~\cite{Gehrmann:2005pd, Gehrmann:2006wg, Heinrich:2007at,
  Heinrich:2009be, Lee:2010cga,Lee:2010ik,Lee:2010ug}
and are collected in the appendix of~\cite{Gehrmann:2010ue}. Using
those, we obtain the final expressions for the 3-loop FFs of the
${\cal O}_{\rm BPS}$ and ${\cal O}_{\cal K}$.

\section{Results of the Form Factors}
\label{sec:Res}

Employing the Feynman diagrammatic approach described in the previous section, we have
first confirmed the form factor  
results for the ${\cal O}_{BPS}$ up to 3-loop level presented
in~\cite{vanNeerven:1985ja, Gehrmann:2011xn} and for ${\cal O}_{{\cal K}}$
up to 2-loop~\cite{Nandan:2014oga}.
In the present letter, we present only the $\epsilon$ expanded results
for the ${\cal F}^{{\cal K},(i)}_{\phi}, ~i=1,2,3$ (see Eq.~(\ref{eq:5})). The exact 
results in terms of $d$ and MIs are too long to present
here and can be obtained from the authors.   
In order to demonstrate the subtleties involved in the choice of
regularization scheme, we have expressed them in terms of $\delta_R$ which is unity in
${\overline{DR}}$ scheme and zero in FDH scheme.  
\vspace{-1cm}
\allowdisplaybreaks[4]
\begin{widetext}
\begin{align}
   \mathcal{F}^{ \mathcal{K},(1)}_{\phi} &=
         \frac{1}{\epsilon^{2}}   \Bigg\{
          - 8
          \Bigg\}
       + \frac{1}{\epsilon}   \Bigg\{
          12
          \Bigg\}
        - 12
          + \zeta_2
       + \epsilon   \Bigg\{
           12
          - \frac{7}{3} \zeta_3
          - \frac{3}{2} \zeta_2
          \Bigg\}
       + \epsilon^2   \Bigg\{
          - 12
          + \frac{7}{2} \zeta_3
          + \frac{3}{2} \zeta_2
          + \frac{47}{80} \zeta_2^2
          \Bigg\}
       + \epsilon^3   \Bigg\{
           12
          - \frac{31}{20} \zeta_5
\nonumber\\
&          - \frac{7}{2} \zeta_3
          - \frac{3}{2} \zeta_2
          + \frac{7}{24} \zeta_2 \zeta_3
          - \frac{141}{160} \zeta_2^2
          \Bigg\}
       + \epsilon^4   \Bigg\{
          - 12
          + \frac{93}{40} \zeta_5
          + \frac{7}{2} \zeta_3
          - \frac{49}{144} \zeta_3^2 
         + \frac{3}{2} \zeta_2
          - \frac{7}{16} \zeta_2 \zeta_3
          + \frac{141}{160} \zeta_2^2
          + \frac{949}{4480} \zeta_2^3
          \Bigg\}
%
%       + \epsilon^5   \Bigg\{
%          + 12
%          - \frac{127}{112} \zeta_7
%          - \frac{93}{40} \zeta_5
%          - \frac{7}{2} \zeta_3
%          + \frac{49}{96} \zeta_3^2
%          - \frac{3}{2} \zeta_2
%          + \frac{31}{160} \zeta_2 \zeta_5
%          + \frac{7}{16} \zeta_2 \zeta_3
%          - \frac{141}{160} \zeta_2^2
%          + \frac{329}{1920} \zeta_2^2 \zeta_3
%          - \frac{2847}{8960} \zeta_2^3
%          \Bigg\}
%
%      
%
\nonumber\\
&
   + \delta_{R} \Bigg[
        - 2
       + 2\,\epsilon   
       + \epsilon^2   \Bigg\{
          - 2
          + \frac{1}{4} \zeta_2
          \Bigg\}
       + \epsilon^3   \Bigg\{
           2
          - \frac{7}{12} \zeta_3
          - \frac{1}{4} \zeta_2
          \Bigg\}
       + \epsilon^4   \Bigg\{
          - 2
          + \frac{7}{12} \zeta_3
          + \frac{1}{4} \zeta_2
          + \frac{47}{320} \zeta_2^2
          \Bigg\}\Bigg],
%
%       + \epsilon^5   \Bigg\{
%          + 2
%          - \frac{31}{80} \zeta_5
%          - \frac{7}{12} \zeta_3
%          - \frac{1}{4} \zeta_2
%          + \frac{7}{96} \zeta_2 \zeta_3
%          - \frac{47}{320} \zeta_2^2
%          \Bigg\}
%
%      
%
\nonumber\\
 \mathcal{F}^{ \mathcal{K},(2)}_{\phi} &= 
        \frac{1}{\epsilon^{4}}   \Bigg\{
           32
          \Bigg\}
       + \frac{1}{\epsilon^{3}}   \Bigg\{
          - 96
          \Bigg\}
       + \frac{1}{\epsilon^{2}}   \Bigg\{
           168
          - 4 \zeta_2
          \Bigg\}
       + \frac{1}{\epsilon}   \Bigg\{
          - 276
          + \frac{50}{3} \zeta_3
          + 24 \zeta_2
          \Bigg\}
        + 438
          - 56 \zeta_3
          - 66 \zeta_2
          - \frac{21}{5} \zeta_2^2
 \nonumber\\
 &
       + \epsilon   \Bigg\{
          - 681
          - \frac{71}{10} \zeta_5   
          + 128 \zeta_3
          + 141 \zeta_2
          - \frac{23}{6} \zeta_2 \zeta_3
          + 15 \zeta_2^2
          \Bigg\}
       + \epsilon^2   \Bigg\{
           \frac{2091}{2}
          + \frac{84}{5} \zeta_5
          - 314 \zeta_3
          + \frac{901}{36} \zeta_3^2
          - \frac{519}{2} \zeta_2                  
 \nonumber\\
 &
          + 26 \zeta_2 \zeta_3
          - \frac{741}{20} \zeta_2^2          
          + \frac{2313}{280} \zeta_2^3
          \Bigg\}
%
%       + \epsilon^3   \Bigg\{
%          - \frac{6369}{4}
%          - \frac{3169}{28} \zeta_7
%          - \frac{12}{5} \zeta_5
%          + 722 \zeta_3
%          - \frac{247}{3} \zeta_3^2
%          + \frac{1761}{4} \zeta_2
%          + \frac{313}{40} \zeta_2 \zeta_5
%          - 82 \zeta_2 \zeta_3
%          + \frac{3813}{40} \zeta_2^2
%          - \frac{1291}{80} \zeta_2^2 \zeta_3
%          - \frac{3891}{140} \zeta_2^3
%          \Bigg\}
%
%       + \epsilon^4   \Bigg\{
%          + \frac{19299}{8}
%          + \frac{18825}{56} \zeta_7
%          - \frac{1227}{10} \zeta_5
%          - \frac{33}{2} \zeta_53
%          - \frac{3085}{2} \zeta_3
%          + \frac{845}{24} \zeta_3 \zeta_5
%          + \frac{577}{3} \zeta_3^2
%          - \frac{5703}{8} \zeta_2
%          - \frac{21}{5} \zeta_2 \zeta_5
%          + \frac{409}{2} \zeta_2 \zeta_3
%          - \frac{1547}{144} \zeta_2 \zeta_3^2
%          - \frac{17931}{80} \zeta_2^2
%          + \frac{1063}{20} \zeta_2^2 \zeta_3
%          + \frac{31401}{560} \zeta_2^3
%          + \frac{50419}{1600} \zeta_2^4
%          \Bigg\}
%
%       
%
   + \delta_{R} \Bigg[
        \frac{1}{\epsilon^{2}}   \Bigg\{
           16
          \Bigg\}
       + \frac{1}{\epsilon}   \Bigg\{
          - 28
          \Bigg\}
        + 46
          - 4 \zeta_2
       + \epsilon   \Bigg\{
          - 73
          + \frac{28}{3} \zeta_3
          + 11 \zeta_2
          \Bigg\}
 \nonumber\\
 &
       + \epsilon^2   \Bigg\{
           \frac{227}{2}
          - \frac{64}{3} \zeta_3
          - \frac{47}{2} \zeta_2
          - \frac{5}{2} \zeta_2^2
          \Bigg\}\Bigg],
%
%       + \epsilon^3   \Bigg\{
%          - \frac{697}{4}
%          - \frac{14}{5} \zeta_5
%          + \frac{157}{3} \zeta_3
%          + \frac{173}{4} \zeta_2
%          - \frac{13}{3} \zeta_2 \zeta_3
%          + \frac{247}{40} \zeta_2^2
%          \Bigg\}
%
%       + \epsilon^4   \Bigg\{
%          + \frac{2123}{8}
%          + \frac{2}{5} \zeta_5
%          - \frac{361}{3} \zeta_3
%          + \frac{247}{18} \zeta_3^2
%          - \frac{587}{8} \zeta_2
%          + \frac{41}{3} \zeta_2 \zeta_3
%          - \frac{1271}{80} \zeta_2^2
%          + \frac{1297}{280} \zeta_2^3
%          \Bigg\}
%
 %
\nonumber\\   
   \mathcal{F}^{ \mathcal{K},(3)}_{\phi} &=
        \frac{1}{\epsilon^{6}}   \Bigg\{
          - \frac{256}{3}
          \Bigg\}
       + \frac{1}{\epsilon^{5}}   \Bigg\{
           384
          \Bigg\}
       + \frac{1}{\epsilon^{4}}   \Bigg\{
          - 960
          \Bigg\}
       + \frac{1}{\epsilon^{3}}   \Bigg\{
           2112
          - \frac{176}{3} \zeta_3
          - 96 \zeta_2
          \Bigg\}
       + \frac{1}{\epsilon^{2}}   \Bigg\{
          - 4368
          + 312 \zeta_3
  \nonumber\\
  &
          + 504 \zeta_2
          + \frac{494}{45} \zeta_2^2
          \Bigg\}
       + \frac{1}{\epsilon}   \Bigg\{
           8760
          + \frac{1756}{15} \zeta_5
          - 1056 \zeta_3
          - 1608 \zeta_2
          + \frac{170}{9} \zeta_2 \zeta_3
          - \frac{459}{5} \zeta_2^2
          \Bigg\}
       - 17316
          - \frac{1014}{5} \zeta_5
 \nonumber\\
 &
          + 3192 \zeta_3
          - \frac{1766}{9} \zeta_3^2
          + 4158 \zeta_2
          - 195 \zeta_2 \zeta_3
          + \frac{3789}{10} \zeta_2^2
          - \frac{22523}{270} \zeta_2^3
   + \delta_{R} \Bigg[
        \frac{1}{\epsilon^{4}}   \Bigg\{
          - 64
          \Bigg\}
       + \frac{1}{\epsilon^{3}}   \Bigg\{
           160
          \Bigg\}
 \nonumber\\
 &
       + \frac{1}{\epsilon^{2}}   \Bigg\{
          - 352
          + 16 \zeta_2
          \Bigg\}
       + \frac{1}{\epsilon}   \Bigg\{
           728
          - 52 \zeta_3
          - 84 \zeta_2
          \Bigg\}
       - 1460
          + 176 \zeta_3
          + 268 \zeta_2
          + \frac{153}{10} \zeta_2^2
          \Bigg],
\end{align}

\end{widetext}
where $\zeta_2=\pi^2/6, \zeta_3\approx1.2020569, \zeta_5\approx1.0369277,
\zeta_7\approx1.0083492.$
The presence of the non-zero coefficients of $\delta_{R}$ signifies the
shortcoming of the FDH scheme in case of Konishi operator. 
We observe that our results for $\delta {\cal F}_{\phi}^{{\cal
    K},(i)}, i=1,2,3$ expressed  
in terms of $d$ and 
MIs contain an overall factor $(6-\delta_R \epsilon)/6$ 
explaining the necessity of correcting the results computed in FDH
scheme by this factor advocated in~\cite{Nandan:2014oga}.

\section{Operator Renormalization}
\label{sec:OperRenorm}

Though the ${\cal N}=4$ SYM is UV finite i.e. neither coupling constant nor
wave function renormalization is required, nevertheless the FFs of the
composite unprotected operators, like Konishi, do involve divergences of the UV
source which are captured by the presence of non-zero UV anomalous
dimensions, $\gamma^{\rho}$. As a consequence, to get rid of the UV
divergences, the FFs are required to undergo UV
renormalization which is performed through the multiplication of an
overall operator renormalization, $Z^{\rho}\left( a,\mu, \epsilon \right)$:
\begin{align}
\label{eq:9}
&\frac{d}{d\ln\mu^2} \ln Z^{\rho} =
  \gamma^{\rho} = \sum\limits_{i=1}^{\infty} a^{i}
  \gamma^{\rho}_{i}\,.
\end{align}
Since $\hat a_s = a_s (\mu_0/\mu)^{\epsilon}$, the solution to the above equation takes
the simple form:
\begin{equation}
\label{eq:17}
Z^{\rho}  = \exp \Big(\sum_{n=1}^{\infty}a^n\frac{2\gamma^\rho_n}{n\epsilon}\Big).
%&= 1 + a \l( \frac{2}{\epsilon} \gamma^{\rho}_{1} \r) + a^{2}
 % \l(\frac{2}{\epsilon^2} \l(\gamma^{\rho}_{1}\r)^{2} +
 % \frac{1}{\epsilon} \gamma^{\rho}_{2} \r)
%\nonumber\\
%&+ a^{3} \l( \frac{4}{3\epsilon^3} \l(\gamma^{\rho}_{1}\r)^{3} +
 % \frac{2}{\epsilon^2} \gamma^{\rho}_1 \gamma^{\rho}_{2} +
  %\frac{2}{3\epsilon} \gamma^{\rho}_{3}\r)\,.
\end{equation}
The UV finite Konishi FFs is obtained as $\left[{\cal F}^{\cal K}_f\right]_{R}
= Z^{\cal K} {\cal F}^{\cal K}_f$, whereas $\left[{\cal F}^{\rm BPS}_f\right]_{R}
= {\cal F}^{\rm BPS}_f$. Since, this is a property of the
associated composite operator, the 
$\gamma^{\rho}$ and so $Z^{\rho}$ are independent of the type as well
as number of the external on-shell states. In the next section, we will discuss the
methodology to obtain the $\gamma^{\rho}$ for the Konishi type of
operators in addition to discussing the IR singularities of the FFs.

\section{Universality of the Pole Structures}

The FFs in ${\cal N}=4$ SYM contain divergences arising from the IR
region which show up as poles in $\epsilon$. The associated pole
structures can be revealed and studied in an elegant way through the
KG-equation~\cite{Sudakov:1954sw, Mueller:1979ih, Collins:1980ih,
  Sen:1981sd} which is obeyed by the 
FFs as a consequence of factorization, gauge and renormalization group
invariances: 
\begin{align}
\label{eq:10}
\frac{d}{d\ln Q^2} \ln {\cal F}^{\rho}_{f}
  = \frac{1}{2} \left[ K^{\rho}_{f} + G^{\rho}_{f} \right]\,.
\end{align}
The $Q^{2}$ independent $K^{\rho}_{f}\left(a, \epsilon \right)$
contains all the poles in $\epsilon$, whereas $G^{\rho}_f \left(a,
  Q^2/\mu^2,\epsilon\right)$ 
involves only the finite terms in $\epsilon \rightarrow 0$. Inspired
from QCD~\cite{Ravindran:2005vv,Bern:2005iz,Ravindran:2006cg}, we propose the
general solution to be
\begin{align}
\label{eq:11}
 \ln {\cal F}^{\rho}_{f}(a, Q^2, \mu^2, \epsilon) =
  \sum_{j=1}^{\infty} {a}^{j} \left(\frac{Q^{2}}{\mu^{2}}\right)^{j
  \frac{\epsilon}{2}} {\cal L}_{f,j}^{\rho}(\epsilon)
\end{align}
with
\begin{align}
\label{eq:12}
{\cal L}_{f,j}^{\rho}(\ep) =&                { \frac{1}{\ep^2} }
                                            \left\{- { \frac{2}{j^{2}} }
                                            A_{j} \right\} 
                                            + { \frac{1}{\ep}
                                            } \left\{  { \frac{1}{j} }
                                            G^{\rho}_{f,j}(\ep)\right\}
\end{align}
where, $A=\sum_{j=1}^{\infty} a^{j} A_{j}$ are the cusp
anomalous dimensions in ${\cal N}=4$ 
SYM. The absence of the superscript $\rho$ and subscript $f$ signifies the
independence of these quantities on the nature of composite operators
as well as external particles. These are determined by looking at the
highest poles of the $\ln F^{\rho}_{f}$ which are found to be
\begin{align}
\label{eq:13}
A_{1} = 4\,,
A_{2} = -8 \zeta_{2}\,,
A_{3} = \frac{176}{5} \zeta_{2}^2
\end{align}
up to 3-loops which are consistent with the results presented in~\cite{Korchemsky:1987wg,Correa:2012nk}.
These are basically the highest transcendental parts of those of
QCD~\cite{Moch:2004pa, Moch:2005tm, Vogt:2004mw,
  Vogt:2000ci}. The other quantities in Eq.~(\ref{eq:12}),
$G^{\rho}_{f,j}$ are postulated, like QCD~\cite{Ravindran:2004mb,
  Moch:2005tm}, to satisfy 
\begin{align}
\label{eq:14}
G^{\rho}_{f,j} (\ep) = 2 \left(B_{j} -
  \gamma^{\rho}_{j}\right)  + f_{j} +
  \sum_{k=1}^{\infty} \epsilon^k g^{\rho,k}_{f,j} 
\end{align}
where, $B=\sum_{j=1}^{\infty} a^{j} B_{j}$ and $f=\sum_{j=1}^{\infty}
a^{j} f_{j}$ are the collinear and soft anomalous
dimensions in ${\cal N}=4$ SYM which are independent of the operators
as well as external legs. For the ${\cal O}_{\rm BPS}$ and ${\cal
  O}_{\cal K}$, we obtain
\begin{align}
\label{eq:15}
&\gamma^{\rm BPS}_{j}=0\,,
\nonumber\\
&\gamma^{\cal K}_{1}= -6\,, \gamma^{\cal K}_{2} = 24\,,
  \gamma^{\cal K}_{3} = -168
\end{align}
up to 3-loop. For the Konishi operator, the results up to 2-loop are
in agreement with the existing ones~\cite{Anselmi:1996mq, Eden:2000mv, Bianchi:2000hn} and
the 3-loop result also matches with previous computations ~\cite{Kotikov:2004er, Eden:2004ua} . By subtracting out the $\gamma_{j}$,
we can only calculate the combination of $(2 B_j+f_j)$. However,
by looking at the similarities between $A_j$ of QCD and ${\cal N}=4$,
we propose
\begin{align}
\label{eq:16}
&B_{1} = 0\,,
B_{2} = 12 \zeta_{3}\,,
B_{3} = 16 \l(-\zeta_{2} \zeta_{3} - 5 \zeta_{5}\r)\,,
\nonumber\\
&f_{1} = 0\,,
f_{2} = -28 \zeta_{3}\,,
f_{3} = \l(\frac{176}{3} \zeta_{2} \zeta_{3} + 192 \zeta_{5} \r)
\end{align}
which are essentially the highest transcendental parts of those of
QCD~\cite{Vogt:2004mw, Moch:2005tm, Ravindran:2004mb}. The other
process dependent constants, that are relevant up to 3-loop, in
Eq.~\~(\ref{eq:14}) are obtained as 
\begin{align}
\label{eq:18}
&g^{{\rm BPS},1}_{\phi,1} = \zeta_2\,,
g^{{\rm BPS},2}_{\phi,1} = - \frac{7}{3} \zeta_3\,,
g^{{\rm BPS},3}_{\phi,1} =\frac{47}{80}\, \zeta_2^2 \,,
\nonumber\\
&g^{{\rm BPS},4}_{\phi,1} =  \frac{7}{24}\zeta_{2} \zeta_{3} -
  \frac{31}{20} \zeta_{5}\,,
g^{{\rm BPS},5}_{\phi,1} = \frac{949}{4480}\zeta_{2}^3 - \frac{49}{144} \zeta_{3}^2\,,
\nonumber\\
&g^{{\rm BPS},1}_{\phi,2} = 0\,,
g^{{\rm BPS},2}_{\phi,2} =  \frac{5}{3} \zeta_2 \zeta_3 - 39
  \zeta_5\,,
g^{{\rm BPS},3}_{\phi,2} = \frac{2623}{140} \zeta_{2}^3 
\nonumber\\
&\quad \quad+ \frac{235}{6}\zeta_{3}^2\,,
g^{{\rm BPS},1}_{\phi,3} = -\frac{12352}{315} \zeta_2^3 -
  \frac{104}{3} \zeta_3^2
\end{align}
for ${\cal O}_{\rm BPS}$. Similarly for the ${\cal O}_{\cal K}$, we
get
%
% \begin{align}
% \label{eq:19}
% &g^{{\cal K},1}_{\phi,1} = -14 + \zeta_{2}\,,
% g^{{\cal K},2}_{\phi,1} = 14 - \frac{3}{2}\zeta_{2} - \frac{7}{3}\zeta_{3}\,,
% \nonumber\\
% &g^{{\cal K},3}_{\phi,1} = -14\, + \frac{7}{4}\,\zeta_2 +
%   \frac{47}{80}\,\zeta_2^2 + \frac{7}{2}\,\zeta_3\,, 
% \nonumber\\
% &g^{{\cal K},4}_{\phi,1} = 14 - \frac{7}{4}\zeta_{2} - \frac{141}{160}
%   \zeta_{2}^2 - \frac{49}{12} \zeta_{3} + \frac{7}{24} \zeta_{2}
%   \zeta_{3} - \frac{31}{20} \zeta_{5}\,,
% \nonumber\\
% &g^{{\cal K},5}_{\phi,1} = -14 + \frac{7}{4} \zeta_{2} +
%   \frac{329}{320} \zeta_{2}^2 + \frac{949}{4480} \zeta_{2}^3 +
%   \frac{49}{12} \zeta_{3} -  \frac{7}{16} \zeta_{2} \zeta_{3} 
% \nonumber\\
% &\quad\quad-
%   \frac{49}{144} \zeta_{3}^2 + \frac{93}{40} \zeta_{5}\,, 
% g^{{\cal K},1}_{\phi,2} = 212 - 48 \zeta_{2}\,,
% \nonumber\\
% &g^{{\cal K},2}_{\phi,2} = -556 + 164 \zeta_{2} +
%   \frac{24}{5}\zeta_{2}^2 + 60 \zeta_{3} +
%   \frac{5}{3}\zeta_{2} \zeta_{3} - 39 \zeta_{5}\,,
% \nonumber\\
% &g^{{\cal K},3}_{\phi,2} = 1170 - 377 \zeta_{2} -
%   \frac{154}{5}\zeta_{2}^2 + \frac{2623}{140} \zeta_{2}^3 - 344 \zeta_{3} + 
%   24 \zeta_{2} \zeta_{3} 
% \nonumber\\
% &\quad\quad+ \frac{235}{6}\zeta_{3}^2 + 108 \zeta_{5}\,,
% g^{{\cal K},1}_{\phi,3} = -2936 - \frac{12352}{315}\zeta_2^3 + 504
%   \zeta_{2} 
% \nonumber\\
% &\quad\quad+
%   \frac{1224}{5} \zeta_{2}^2 - 648 \zeta_{3} 
% - \frac{104}{3} \zeta_{3}^2 +
%   720 \zeta_{5}\,.
% \end{align}
%
\begin{align}
\label{eq:19}
&g^{{\cal K},1}_{\phi,1} = g^{{\rm BPS},1}_{\phi,1} - 14 \,,
g^{{\cal K},2}_{\phi,1} = g^{{\rm BPS},2}_{\phi,1} + 14 - \frac{3}{2}\zeta_{2}\,,
\nonumber\\
&g^{{\cal K},3}_{\phi,1} = g^{{\rm BPS},3}_{\phi,1} -14 + \frac{7}{4}\zeta_2 + \frac{7}{2}\zeta_3\,, 
g^{{\cal K},4}_{\phi,1} = g^{{\rm BPS},4}_{\phi,1} + 14 
\nonumber\\
&\quad\quad- \frac{7}{4}\zeta_{2} - \frac{141}{160}
  \zeta_{2}^2 - \frac{49}{12} \zeta_{3} \,,
g^{{\cal K},5}_{\phi,1} =g^{{\rm BPS},5}_{\phi,1} - 14 + \frac{7}{4}
  \zeta_{2} 
\nonumber\\
&\quad\quad+
  \frac{329}{320} \zeta_{2}^2 +
  \frac{49}{12} \zeta_{3} -  \frac{7}{16} \zeta_{2} \zeta_{3} 
+ \frac{93}{40} \zeta_{5}\,, 
\nonumber\\
&
g^{{\cal K},1}_{\phi,2} =g^{{\rm BPS},1}_{\phi,2} + 212 - 48 \zeta_{2}\,,
g^{{\cal K},2}_{\phi,2} = g^{{\rm BPS},2}_{\phi,2} -556 + 164
  \zeta_{2} 
\nonumber\\
&\quad\quad+
  \frac{24}{5}\zeta_{2}^2 + 60 \zeta_{3} \,,
g^{{\cal K},3}_{\phi,2} = g^{{\rm BPS},3}_{\phi,2} + 1170 - 377
  \zeta_{2} 
\nonumber\\
&\quad \quad-
  \frac{154}{5}\zeta_{2}^2  - 344 \zeta_{3} + 
  24 \zeta_{2} \zeta_{3} + 108 \zeta_{5}\,,
g^{{\cal K},1}_{\phi,3} = g^{{\rm BPS},1}_{\phi,3} 
\nonumber\\
&\quad \quad-2936  + 504
  \zeta_{2} 
+
  \frac{1224}{5} \zeta_{2}^2 - 648 \zeta_{3} + 720 \zeta_{5}\,.
\end{align}
In a clear contrast to that of QCD, due to absence of
the non-zero $\beta$-functions in ${\cal N}=4$ SYM, all the higher
poles vanish in Eq.~(\ref{eq:12}). We observe that the leading transcendental terms 
in the operator dependent parts of the FFs of ${\cal O}_{\cal K}$ and
${\cal O}_{\rm BPS}$, namely $g_{\phi,j}^{\rho,k}$, coincide. This
is indeed the case with QCD form factors when the color factors are chosen suitably.
%{\color{red}Using the results of the above quantities, the ${\cal
%  F}^{\rho}_{f}$ up to 3-loop level can be determined.}

\vspace{0.4cm}
\section{Form Factors Beyond Three Loop}
\vspace{-0.3cm}
The KG equation~(\ref{eq:10}) enables us to predict all the poles but 
constant term of the FFs at 4-loop. Expanding the results of the FFs
of previous orders sufficiently high, using the $A_{4}$ ~\cite{Bern:2005iz, Bern:2006ew, Cachazo:2006az, Henn:2010ir,   Henn:2013wfa}, $\l(2B_{4}+f_{4}\r)$ ~\cite{Beisert:2006ez, Cachazo:2007ad,Henn:2013wfa}\,\,denoted by $\alpha$ from~\cite{Bern:2008ap} and $\gamma^{\cal K}_{4}$
from~\cite{Fiamberti:2007rj, Fiamberti:2008sh, Velizhanin:2008jd, Velizhanin:2009gv} we obtain ${\cal F}^{{\cal K},(4)}_{\phi}|_{\rm poles} $:
\vspace{-1cm}
\allowdisplaybreaks[4]
\begin{widetext}
\begin{align}
\label{eq:20}
{\cal F}^{{\cal K},(4)}_{\phi}|_{\rm poles} &= \frac{1}{\epsilon^{8}}   \Bigg\{
           \frac{512}{3}
          \Bigg\}
       + \frac{1}{\epsilon^{7}}   \Bigg\{
          - 1024
          \Bigg\}
       + \frac{1}{\epsilon^{6}}   \Bigg\{
           \frac{10496}{3}
          + \frac{128}{3} \zeta_2
          \Bigg\}
       + \frac{1}{\epsilon^{5}}   \Bigg\{
          - \frac{28928}{3}
          + \frac{1216}{9} \zeta_3
          + 128 \zeta_2
          \Bigg\}
       + \frac{1}{\epsilon^{4}}   \Bigg\{
           \frac{72992}{3}
\nonumber\\
&          - \frac{3008}{3} \zeta_3
          - \frac{5344}{3} \zeta_2
          + \frac{40}{9} \zeta_2^2
          \Bigg\}
       + \frac{1}{\epsilon^{3}}   \Bigg\{
          - \frac{176192}{3}
          - \frac{8656}{15} \zeta_5
          + \frac{42064}{9} \zeta_3
          + \frac{25024}{3} \zeta_2
          - \frac{184}{3} \zeta_2 \zeta_3
          + \frac{4256}{15} \zeta_2^2
          \Bigg\}
 \nonumber\\
 &
       + \frac{1}{\epsilon^{2}}   \Bigg\{
           \frac{416096}{3}
          + \frac{5072}{5} \zeta_5
          - \frac{151648}{9} \zeta_3
          + \frac{21706}{27} \zeta_3^2
          - \frac{85408}{3} \zeta_2
          + 736 \zeta_2 \zeta_3
          - \frac{18488}{9} \zeta_2^2
          + \frac{381908}{945} \zeta_2^3
          \Bigg\}
\nonumber\\
&
       + \frac{1}{\epsilon}   \Bigg\{
          - \frac{973136}{3}
          - 4 \alpha
          - \frac{536894}{63} \zeta_7
          + \frac{160412}{15} \zeta_5
          + \frac{409192}{9} \zeta_3
          - \frac{18680}{9} \zeta_3^2
          + \frac{254536}{3} \zeta_2
          + \frac{33938}{45} \zeta_2 \zeta_5
 \nonumber\\
&         - \frac{14336}{3} \zeta_2 \zeta_3
          + \frac{67664}{9} \zeta_2^2
          - \frac{14590}{27} \zeta_2^2 \zeta_3
          - \frac{333712}{315} \zeta_2^3
          \Bigg\}\,,
\end{align}
\end{widetext}
where $\alpha = -(77.56 \pm 0.02) $.
Explicit computation is required to get the constant terms. The exact
matching of the highest transcendental terms of ${\cal O}_{\cal K}$
and ${\cal O}_{\rm BPS}$ at 4-loop holds true, similar to the
previous orders.

To summarize, we have presented for the first time the third order
corrections to the on-shell form factor of  the
Konishi operator 
employing the standard Feynman diagrammatic approach.  The computation
is performed in the $\overline {DR}$ 
and FDH schemes in order to demonstrate the subtleties involved with
the latter one when applied to
composite operators that depend on the space-time dimension $d$.  We have
shown up to third order, the results for  
d-independent operators are insensitive to the regularization schemes, while for the d-dependent
operators, results in  
FDH scheme need to be corrected by suitable $d$ dependent terms in order to preserve 
the SUSY.  
It is also demonstrated that the FFs of Konishi operator computed {\it only} in ${\overline{DR}}$ 
satisfies $KG$ equation and also can be described in terms of universal cusp, collinear 
and soft anomalous dimensions.  This implies that infrared
factorization of FFs in ${\cal N}=4$ SYM theory  
can be established only if the supersymmetric preserving
regularisation is used when computing higher 
order effects. Up to third order, we find that the anomalous dimensions resulting from IR region
are related to those of QCD when the color factors are adjusted suitably. In addition, we 
confirm the UV anomalous dimensions of the Konishi operator up to third order, whose extraction
depends on the universal IR structure of the FFs.  This provides a consistency check
of both the UV and IR structure of FFs in ${\cal N}=4$.  
Agreements of our 3-loop result for the FFs of ${\cal O}_{BPS}$ and 2-loop result
for the FFs of ${\cal O}_{\cal K}$ computed using Feynman diagrammatic 
techniques with those obtained using on-shell methods
in~\cite{vanNeerven:1985ja, Gehrmann:2011xn} and~\cite{Nandan:2014oga}, respectively,
establish the power and reliability of various state-of-the-arts
approaches to deal with higher order corrections in QFT.   Finally,  we  use KG equation
to predict four loop results for both BPS and Konishi operators up to $\epsilon^{-1}$.
 
We would like to thank T.\,Gehrmann, J.\,Henn,  R.\,N.\,Lee, M.\,Mahakhud, P.\,Nogueira  and A.\,Tripathi for useful discussions. We would like to thank D.\,Jatkar and A.\,Sen for carefully going over the manuscript and providing useful suggestions.

\bibliography{konishi} 

\begin{thebibliography}{61}%
\makeatletter
\providecommand \@ifxundefined [1]{%
 \@ifx{#1\undefined}
}%
\providecommand \@ifnum [1]{%
 \ifnum #1\expandafter \@firstoftwo
 \else \expandafter \@secondoftwo
 \fi
}%
\providecommand \@ifx [1]{%
 \ifx #1\expandafter \@firstoftwo
 \else \expandafter \@secondoftwo
 \fi
}%
\providecommand \natexlab [1]{#1}%
\providecommand \enquote  [1]{``#1''}%
\providecommand \bibnamefont  [1]{#1}%
\providecommand \bibfnamefont [1]{#1}%
\providecommand \citenamefont [1]{#1}%
\providecommand \href@noop [0]{\@secondoftwo}%
\providecommand \href [0]{\begingroup \@sanitize@url \@href}%
\providecommand \@href[1]{\@@startlink{#1}\@@href}%
\providecommand \@@href[1]{\endgroup#1\@@endlink}%
\providecommand \@sanitize@url [0]{\catcode `\\12\catcode `\$12\catcode
  `\&12\catcode `\#12\catcode `\^12\catcode `\_12\catcode `\%12\relax}%
\providecommand \@@startlink[1]{}%
\providecommand \@@endlink[0]{}%
\providecommand \url  [0]{\begingroup\@sanitize@url \@url }%
\providecommand \@url [1]{\endgroup\@href {#1}{\urlprefix }}%
\providecommand \urlprefix  [0]{URL }%
\providecommand \Eprint [0]{\href }%
\providecommand \doibase [0]{http://dx.doi.org/}%
\providecommand \selectlanguage [0]{\@gobble}%
\providecommand \bibinfo  [0]{\@secondoftwo}%
\providecommand \bibfield  [0]{\@secondoftwo}%
\providecommand \translation [1]{[#1]}%
\providecommand \BibitemOpen [0]{}%
\providecommand \bibitemStop [0]{}%
\providecommand \bibitemNoStop [0]{.\EOS\space}%
\providecommand \EOS [0]{\spacefactor3000\relax}%
\providecommand \BibitemShut  [1]{\csname bibitem#1\endcsname}%
\let\auto@bib@innerbib\@empty
%</preamble>
\bibitem [{\citenamefont {Brink}\ \emph {et~al.}(1977)\citenamefont {Brink},
  \citenamefont {Schwarz},\ and\ \citenamefont {Scherk}}]{Brink:1976bc}%
  \BibitemOpen
  \bibfield  {author} {\bibinfo {author} {\bibfnamefont {L.}~\bibnamefont
  {Brink}}, \bibinfo {author} {\bibfnamefont {J.~H.}\ \bibnamefont {Schwarz}},
  \ and\ \bibinfo {author} {\bibfnamefont {J.}~\bibnamefont {Scherk}},\ }\href
  {\doibase 10.1016/0550-3213(77)90328-5} {\bibfield  {journal} {\bibinfo
  {journal} {Nucl. Phys.}\ }\textbf {\bibinfo {volume} {B121}},\ \bibinfo
  {pages} {77} (\bibinfo {year} {1977})}\BibitemShut {NoStop}%
%%CITATION = NUPHA,B121,77;%%
\bibitem [{\citenamefont {Gliozzi}\ \emph {et~al.}(1977)\citenamefont
  {Gliozzi}, \citenamefont {Scherk},\ and\ \citenamefont
  {Olive}}]{Gliozzi:1976qd}%
  \BibitemOpen
  \bibfield  {author} {\bibinfo {author} {\bibfnamefont {F.}~\bibnamefont
  {Gliozzi}}, \bibinfo {author} {\bibfnamefont {J.}~\bibnamefont {Scherk}}, \
  and\ \bibinfo {author} {\bibfnamefont {D.~I.}\ \bibnamefont {Olive}},\ }\href
  {\doibase 10.1016/0550-3213(77)90206-1} {\bibfield  {journal} {\bibinfo
  {journal} {Nucl. Phys.}\ }\textbf {\bibinfo {volume} {B122}},\ \bibinfo
  {pages} {253} (\bibinfo {year} {1977})}\BibitemShut {NoStop}%
%%CITATION = NUPHA,B122,253;%%
\bibitem [{\citenamefont {van Neerven}(1986)}]{vanNeerven:1985ja}%
  \BibitemOpen
  \bibfield  {author} {\bibinfo {author} {\bibfnamefont {W.~L.}\ \bibnamefont
  {van Neerven}},\ }\href {\doibase 10.1007/BF01571808} {\bibfield  {journal}
  {\bibinfo  {journal} {Z. Phys.}\ }\textbf {\bibinfo {volume} {C30}},\
  \bibinfo {pages} {595} (\bibinfo {year} {1986})}\BibitemShut {NoStop}%
%%CITATION = ZEPYA,C30,595;%%
\bibitem [{\citenamefont {Gehrmann}\ \emph {et~al.}(2012)\citenamefont
  {Gehrmann}, \citenamefont {Henn},\ and\ \citenamefont
  {Huber}}]{Gehrmann:2011xn}%
  \BibitemOpen
  \bibfield  {author} {\bibinfo {author} {\bibfnamefont {T.}~\bibnamefont
  {Gehrmann}}, \bibinfo {author} {\bibfnamefont {J.~M.}\ \bibnamefont {Henn}},
  \ and\ \bibinfo {author} {\bibfnamefont {T.}~\bibnamefont {Huber}},\ }\href
  {\doibase 10.1007/JHEP03(2012)101} {\bibfield  {journal} {\bibinfo  {journal}
  {JHEP}\ }\textbf {\bibinfo {volume} {03}},\ \bibinfo {pages} {101} (\bibinfo
  {year} {2012})}\BibitemShut {NoStop}%
\bibitem [{\citenamefont {Nandan}\ \emph {et~al.}(2015)\citenamefont {Nandan},
  \citenamefont {Sieg}, \citenamefont {Wilhelm},\ and\ \citenamefont
  {Yang}}]{Nandan:2014oga}%
  \BibitemOpen
  \bibfield  {author} {\bibinfo {author} {\bibfnamefont {D.}~\bibnamefont
  {Nandan}}, \bibinfo {author} {\bibfnamefont {C.}~\bibnamefont {Sieg}},
  \bibinfo {author} {\bibfnamefont {M.}~\bibnamefont {Wilhelm}}, \ and\
  \bibinfo {author} {\bibfnamefont {G.}~\bibnamefont {Yang}},\ }\href {\doibase
  10.1007/JHEP06(2015)156} {\bibfield  {journal} {\bibinfo  {journal} {JHEP}\
  }\textbf {\bibinfo {volume} {06}},\ \bibinfo {pages} {156} (\bibinfo {year}
  {2015})}\BibitemShut {NoStop}%
\bibitem [{\citenamefont {Sudakov}(1956)}]{Sudakov:1954sw}%
  \BibitemOpen
  \bibfield  {author} {\bibinfo {author} {\bibfnamefont {V.~V.}\ \bibnamefont
  {Sudakov}},\ }\href@noop {} {\bibfield  {journal} {\bibinfo  {journal} {Sov.
  Phys. JETP}\ }\textbf {\bibinfo {volume} {3}},\ \bibinfo {pages} {65}
  (\bibinfo {year} {1956})},\ \bibinfo {note} {[Zh. Eksp. Teor.
  Fiz.30,87(1956)]}\BibitemShut {NoStop}%
%%CITATION = SPHJA,3,65;%%
\bibitem [{\citenamefont {Mueller}(1979)}]{Mueller:1979ih}%
  \BibitemOpen
  \bibfield  {author} {\bibinfo {author} {\bibfnamefont {A.~H.}\ \bibnamefont
  {Mueller}},\ }\href {\doibase 10.1103/PhysRevD.20.2037} {\bibfield  {journal}
  {\bibinfo  {journal} {Phys. Rev.}\ }\textbf {\bibinfo {volume} {D20}},\
  \bibinfo {pages} {2037} (\bibinfo {year} {1979})}\BibitemShut {NoStop}%
%%CITATION = PHRVA,D20,2037;%%
\bibitem [{\citenamefont {Collins}(1980)}]{Collins:1980ih}%
  \BibitemOpen
  \bibfield  {author} {\bibinfo {author} {\bibfnamefont {J.~C.}\ \bibnamefont
  {Collins}},\ }\href {\doibase 10.1103/PhysRevD.22.1478} {\bibfield  {journal}
  {\bibinfo  {journal} {Phys. Rev.}\ }\textbf {\bibinfo {volume} {D22}},\
  \bibinfo {pages} {1478} (\bibinfo {year} {1980})}\BibitemShut {NoStop}%
%%CITATION = PHRVA,D22,1478;%%
\bibitem [{\citenamefont {Sen}(1981)}]{Sen:1981sd}%
  \BibitemOpen
  \bibfield  {author} {\bibinfo {author} {\bibfnamefont {A.}~\bibnamefont
  {Sen}},\ }\href {\doibase 10.1103/PhysRevD.24.3281} {\bibfield  {journal}
  {\bibinfo  {journal} {Phys. Rev.}\ }\textbf {\bibinfo {volume} {D24}},\
  \bibinfo {pages} {3281} (\bibinfo {year} {1981})}\BibitemShut {NoStop}%
%%CITATION = PHRVA,D24,3281;%%
\bibitem [{\citenamefont {Ravindran}\ \emph {et~al.}(2005)\citenamefont
  {Ravindran}, \citenamefont {Smith},\ and\ \citenamefont {van
  Neerven}}]{Ravindran:2004mb}%
  \BibitemOpen
  \bibfield  {author} {\bibinfo {author} {\bibfnamefont {V.}~\bibnamefont
  {Ravindran}}, \bibinfo {author} {\bibfnamefont {J.}~\bibnamefont {Smith}}, \
  and\ \bibinfo {author} {\bibfnamefont {W.~L.}\ \bibnamefont {van Neerven}},\
  }\href {\doibase 10.1016/j.nuclphysb.2004.10.039} {\bibfield  {journal}
  {\bibinfo  {journal} {Nucl. Phys.}\ }\textbf {\bibinfo {volume} {B704}},\
  \bibinfo {pages} {332} (\bibinfo {year} {2005})}\BibitemShut {NoStop}%
\bibitem [{\citenamefont {Moch}\ \emph {et~al.}(2005)\citenamefont {Moch},
  \citenamefont {Vermaseren},\ and\ \citenamefont {Vogt}}]{Moch:2005tm}%
  \BibitemOpen
  \bibfield  {author} {\bibinfo {author} {\bibfnamefont {S.}~\bibnamefont
  {Moch}}, \bibinfo {author} {\bibfnamefont {J.~A.~M.}\ \bibnamefont
  {Vermaseren}}, \ and\ \bibinfo {author} {\bibfnamefont {A.}~\bibnamefont
  {Vogt}},\ }\href {\doibase 10.1016/j.physletb.2005.08.067} {\bibfield
  {journal} {\bibinfo  {journal} {Phys. Lett.}\ }\textbf {\bibinfo {volume}
  {B625}},\ \bibinfo {pages} {245} (\bibinfo {year} {2005})}\BibitemShut
  {NoStop}%
\bibitem [{\citenamefont {Bern}\ \emph {et~al.}(2005)\citenamefont {Bern},
  \citenamefont {Dixon},\ and\ \citenamefont {Smirnov}}]{Bern:2005iz}%
  \BibitemOpen
  \bibfield  {author} {\bibinfo {author} {\bibfnamefont {Z.}~\bibnamefont
  {Bern}}, \bibinfo {author} {\bibfnamefont {L.~J.}\ \bibnamefont {Dixon}}, \
  and\ \bibinfo {author} {\bibfnamefont {V.~A.}\ \bibnamefont {Smirnov}},\
  }\href {\doibase 10.1103/PhysRevD.72.085001} {\bibfield  {journal} {\bibinfo
  {journal} {Phys. Rev.}\ }\textbf {\bibinfo {volume} {D72}},\ \bibinfo {pages}
  {085001} (\bibinfo {year} {2005})}\BibitemShut {NoStop}%
\bibitem [{\citenamefont {Bern}\ and\ \citenamefont
  {Kosower}(1992)}]{Bern:1991aq}%
  \BibitemOpen
  \bibfield  {author} {\bibinfo {author} {\bibfnamefont {Z.}~\bibnamefont
  {Bern}}\ and\ \bibinfo {author} {\bibfnamefont {D.~A.}\ \bibnamefont
  {Kosower}},\ }\href {\doibase 10.1016/0550-3213(92)90134-W} {\bibfield
  {journal} {\bibinfo  {journal} {Nucl. Phys.}\ }\textbf {\bibinfo {volume}
  {B379}},\ \bibinfo {pages} {451} (\bibinfo {year} {1992})}\BibitemShut
  {NoStop}%
%%CITATION = NUPHA,B379,451;%%
\bibitem [{\citenamefont {Bern}\ \emph {et~al.}(2002)\citenamefont {Bern},
  \citenamefont {De~Freitas}, \citenamefont {Dixon},\ and\ \citenamefont
  {Wong}}]{Bern:2002zk}%
  \BibitemOpen
  \bibfield  {author} {\bibinfo {author} {\bibfnamefont {Z.}~\bibnamefont
  {Bern}}, \bibinfo {author} {\bibfnamefont {A.}~\bibnamefont {De~Freitas}},
  \bibinfo {author} {\bibfnamefont {L.~J.}\ \bibnamefont {Dixon}}, \ and\
  \bibinfo {author} {\bibfnamefont {H.~L.}\ \bibnamefont {Wong}},\ }\href
  {\doibase 10.1103/PhysRevD.66.085002} {\bibfield  {journal} {\bibinfo
  {journal} {Phys. Rev.}\ }\textbf {\bibinfo {volume} {D66}},\ \bibinfo {pages}
  {085002} (\bibinfo {year} {2002})}\BibitemShut {NoStop}%
\bibitem [{\citenamefont {Siegel}(1979)}]{Siegel:1979wq}%
  \BibitemOpen
  \bibfield  {author} {\bibinfo {author} {\bibfnamefont {W.}~\bibnamefont
  {Siegel}},\ }\href {\doibase 10.1016/0370-2693(79)90282-X} {\bibfield
  {journal} {\bibinfo  {journal} {Phys. Lett.}\ }\textbf {\bibinfo {volume}
  {B84}},\ \bibinfo {pages} {193} (\bibinfo {year} {1979})}\BibitemShut
  {NoStop}%
%%CITATION = PHLTA,B84,193;%%
\bibitem [{\citenamefont {Capper}\ \emph {et~al.}(1980)\citenamefont {Capper},
  \citenamefont {Jones},\ and\ \citenamefont {van
  Nieuwenhuizen}}]{Capper:1979ns}%
  \BibitemOpen
  \bibfield  {author} {\bibinfo {author} {\bibfnamefont {D.~M.}\ \bibnamefont
  {Capper}}, \bibinfo {author} {\bibfnamefont {D.~R.~T.}\ \bibnamefont
  {Jones}}, \ and\ \bibinfo {author} {\bibfnamefont {P.}~\bibnamefont {van
  Nieuwenhuizen}},\ }\href {\doibase 10.1016/0550-3213(80)90244-8} {\bibfield
  {journal} {\bibinfo  {journal} {Nucl. Phys.}\ }\textbf {\bibinfo {volume}
  {B167}},\ \bibinfo {pages} {479} (\bibinfo {year} {1980})}\BibitemShut
  {NoStop}%
%%CITATION = NUPHA,B167,479;%%
\bibitem [{\citenamefont {Ahmed}\ \emph
  {et~al.}(2015{\natexlab{a}})\citenamefont {Ahmed}, \citenamefont {Das},
  \citenamefont {Mathews}, \citenamefont {Rana},\ and\ \citenamefont
  {Ravindran}}]{Ahmed:2015qia}%
  \BibitemOpen
  \bibfield  {author} {\bibinfo {author} {\bibfnamefont {T.}~\bibnamefont
  {Ahmed}}, \bibinfo {author} {\bibfnamefont {G.}~\bibnamefont {Das}}, \bibinfo
  {author} {\bibfnamefont {P.}~\bibnamefont {Mathews}}, \bibinfo {author}
  {\bibfnamefont {N.}~\bibnamefont {Rana}}, \ and\ \bibinfo {author}
  {\bibfnamefont {V.}~\bibnamefont {Ravindran}},\ }\href {\doibase
  10.1007/JHEP12(2015)084} {\bibfield  {journal} {\bibinfo  {journal} {JHEP}\
  }\textbf {\bibinfo {volume} {12}},\ \bibinfo {pages} {084} (\bibinfo {year}
  {2015}{\natexlab{a}})}\BibitemShut {NoStop}%
\bibitem [{\citenamefont {Ahmed}\ \emph
  {et~al.}(2015{\natexlab{b}})\citenamefont {Ahmed}, \citenamefont {Gehrmann},
  \citenamefont {Mathews}, \citenamefont {Rana},\ and\ \citenamefont
  {Ravindran}}]{Ahmed:2015qpa}%
  \BibitemOpen
  \bibfield  {author} {\bibinfo {author} {\bibfnamefont {T.}~\bibnamefont
  {Ahmed}}, \bibinfo {author} {\bibfnamefont {T.}~\bibnamefont {Gehrmann}},
  \bibinfo {author} {\bibfnamefont {P.}~\bibnamefont {Mathews}}, \bibinfo
  {author} {\bibfnamefont {N.}~\bibnamefont {Rana}}, \ and\ \bibinfo {author}
  {\bibfnamefont {V.}~\bibnamefont {Ravindran}},\ }\href {\doibase
  10.1007/JHEP11(2015)169} {\bibfield  {journal} {\bibinfo  {journal} {JHEP}\
  }\textbf {\bibinfo {volume} {11}},\ \bibinfo {pages} {169} (\bibinfo {year}
  {2015}{\natexlab{b}})}\BibitemShut {NoStop}%
\bibitem [{\citenamefont {Nogueira}(1993)}]{Nogueira:1991ex}%
  \BibitemOpen
  \bibfield  {author} {\bibinfo {author} {\bibfnamefont {P.}~\bibnamefont
  {Nogueira}},\ }\href {\doibase 10.1006/jcph.1993.1074} {\bibfield  {journal}
  {\bibinfo  {journal} {J. Comput. Phys.}\ }\textbf {\bibinfo {volume} {105}},\
  \bibinfo {pages} {279} (\bibinfo {year} {1993})}\BibitemShut {NoStop}%
%%CITATION = JCTPA,105,279;%%
\bibitem [{\citenamefont {Vermaseren}(2000)}]{Vermaseren:2000nd}%
  \BibitemOpen
  \bibfield  {author} {\bibinfo {author} {\bibfnamefont {J.~A.~M.}\
  \bibnamefont {Vermaseren}},\ }\href@noop {} {\  (\bibinfo {year} {2000})},\
  \Eprint {http://arxiv.org/abs/math-ph/0010025} {arXiv:math-ph/0010025
  [math-ph]} \BibitemShut {NoStop}%
%%CITATION = MATH-PH/0010025;%%
\bibitem [{\citenamefont {Tkachov}(1981)}]{Tkachov:1981wb}%
  \BibitemOpen
  \bibfield  {author} {\bibinfo {author} {\bibfnamefont {F.}~\bibnamefont
  {Tkachov}},\ }\href {\doibase 10.1016/0370-2693(81)90288-4} {\bibfield
  {journal} {\bibinfo  {journal} {Phys.Lett.}\ }\textbf {\bibinfo {volume}
  {B100}},\ \bibinfo {pages} {65} (\bibinfo {year} {1981})}\BibitemShut
  {NoStop}%
%%CITATION = PHLTA,B100,65;%%
\bibitem [{\citenamefont {Chetyrkin}\ and\ \citenamefont
  {Tkachov}(1981)}]{Chetyrkin:1981qh}%
  \BibitemOpen
  \bibfield  {author} {\bibinfo {author} {\bibfnamefont {K.}~\bibnamefont
  {Chetyrkin}}\ and\ \bibinfo {author} {\bibfnamefont {F.}~\bibnamefont
  {Tkachov}},\ }\href {\doibase 10.1016/0550-3213(81)90199-1} {\bibfield
  {journal} {\bibinfo  {journal} {Nucl.Phys.}\ }\textbf {\bibinfo {volume}
  {B192}},\ \bibinfo {pages} {159} (\bibinfo {year} {1981})}\BibitemShut
  {NoStop}%
%%CITATION = NUPHA,B192,159;%%
\bibitem [{\citenamefont {Gehrmann}\ and\ \citenamefont
  {Remiddi}(2000)}]{Gehrmann:1999as}%
  \BibitemOpen
  \bibfield  {author} {\bibinfo {author} {\bibfnamefont {T.}~\bibnamefont
  {Gehrmann}}\ and\ \bibinfo {author} {\bibfnamefont {E.}~\bibnamefont
  {Remiddi}},\ }\href {\doibase 10.1016/S0550-3213(00)00223-6} {\bibfield
  {journal} {\bibinfo  {journal} {Nucl.Phys.}\ }\textbf {\bibinfo {volume}
  {B580}},\ \bibinfo {pages} {485} (\bibinfo {year} {2000})}\BibitemShut
  {NoStop}%
\bibitem [{\citenamefont {Lee}(2008)}]{Lee:2008tj}%
  \BibitemOpen
  \bibfield  {author} {\bibinfo {author} {\bibfnamefont {R.~N.}\ \bibnamefont
  {Lee}},\ }\href {\doibase 10.1088/1126-6708/2008/07/031} {\bibfield
  {journal} {\bibinfo  {journal} {JHEP}\ }\textbf {\bibinfo {volume} {07}},\
  \bibinfo {pages} {031} (\bibinfo {year} {2008})}\BibitemShut {NoStop}%
\bibitem [{\citenamefont {Lee}(2012)}]{Lee:2012cn}%
  \BibitemOpen
  \bibfield  {author} {\bibinfo {author} {\bibfnamefont {R.}~\bibnamefont
  {Lee}},\ }\href@noop {} {\  (\bibinfo {year} {2012})},\ \Eprint
  {http://arxiv.org/abs/1212.2685} {arXiv:1212.2685 [hep-ph]} \BibitemShut
  {NoStop}%
%%CITATION = ARXIV:1212.2685;%%
\bibitem [{\citenamefont {Lee}(2014)}]{Lee:2013mka}%
  \BibitemOpen
  \bibfield  {author} {\bibinfo {author} {\bibfnamefont {R.~N.}\ \bibnamefont
  {Lee}},\ }\href {\doibase 10.1088/1742-6596/523/1/012059} {\bibfield
  {journal} {\bibinfo  {journal} {J.Phys.Conf.Ser.}\ }\textbf {\bibinfo
  {volume} {523}},\ \bibinfo {pages} {012059} (\bibinfo {year}
  {2014})}\BibitemShut {NoStop}%
\bibitem [{\citenamefont {Anastasiou}\ and\ \citenamefont
  {Lazopoulos}(2004)}]{Anastasiou:2004vj}%
  \BibitemOpen
  \bibfield  {author} {\bibinfo {author} {\bibfnamefont {C.}~\bibnamefont
  {Anastasiou}}\ and\ \bibinfo {author} {\bibfnamefont {A.}~\bibnamefont
  {Lazopoulos}},\ }\href {\doibase 10.1088/1126-6708/2004/07/046} {\bibfield
  {journal} {\bibinfo  {journal} {JHEP}\ }\textbf {\bibinfo {volume} {07}},\
  \bibinfo {pages} {046} (\bibinfo {year} {2004})}\BibitemShut {NoStop}%
\bibitem [{\citenamefont {Smirnov}(2008)}]{Smirnov:2008iw}%
  \BibitemOpen
  \bibfield  {author} {\bibinfo {author} {\bibfnamefont {A.~V.}\ \bibnamefont
  {Smirnov}},\ }\href {\doibase 10.1088/1126-6708/2008/10/107} {\bibfield
  {journal} {\bibinfo  {journal} {JHEP}\ }\textbf {\bibinfo {volume} {10}},\
  \bibinfo {pages} {107} (\bibinfo {year} {2008})}\BibitemShut {NoStop}%
\bibitem [{\citenamefont {von Manteuffel}\ and\ \citenamefont
  {Studerus}(2012)}]{vonManteuffel:2012np}%
  \BibitemOpen
  \bibfield  {author} {\bibinfo {author} {\bibfnamefont {A.}~\bibnamefont {von
  Manteuffel}}\ and\ \bibinfo {author} {\bibfnamefont {C.}~\bibnamefont
  {Studerus}},\ }\href@noop {} {\  (\bibinfo {year} {2012})},\ \Eprint
  {http://arxiv.org/abs/1201.4330} {arXiv:1201.4330 [hep-ph]} \BibitemShut
  {NoStop}%
%%CITATION = ARXIV:1201.4330;%%
\bibitem [{\citenamefont {Studerus}(2010)}]{Studerus:2009ye}%
  \BibitemOpen
  \bibfield  {author} {\bibinfo {author} {\bibfnamefont {C.}~\bibnamefont
  {Studerus}},\ }\href {\doibase 10.1016/j.cpc.2010.03.012} {\bibfield
  {journal} {\bibinfo  {journal} {Comput. Phys. Commun.}\ }\textbf {\bibinfo
  {volume} {181}},\ \bibinfo {pages} {1293} (\bibinfo {year}
  {2010})}\BibitemShut {NoStop}%
\bibitem [{\citenamefont {Gehrmann}\ \emph {et~al.}(2005)\citenamefont
  {Gehrmann}, \citenamefont {Huber},\ and\ \citenamefont
  {Maitre}}]{Gehrmann:2005pd}%
  \BibitemOpen
  \bibfield  {author} {\bibinfo {author} {\bibfnamefont {T.}~\bibnamefont
  {Gehrmann}}, \bibinfo {author} {\bibfnamefont {T.}~\bibnamefont {Huber}}, \
  and\ \bibinfo {author} {\bibfnamefont {D.}~\bibnamefont {Maitre}},\ }\href
  {\doibase 10.1016/j.physletb.2005.07.019} {\bibfield  {journal} {\bibinfo
  {journal} {Phys. Lett.}\ }\textbf {\bibinfo {volume} {B622}},\ \bibinfo
  {pages} {295} (\bibinfo {year} {2005})}\BibitemShut {NoStop}%
\bibitem [{\citenamefont {Gehrmann}\ \emph {et~al.}(2006)\citenamefont
  {Gehrmann}, \citenamefont {Heinrich}, \citenamefont {Huber},\ and\
  \citenamefont {Studerus}}]{Gehrmann:2006wg}%
  \BibitemOpen
  \bibfield  {author} {\bibinfo {author} {\bibfnamefont {T.}~\bibnamefont
  {Gehrmann}}, \bibinfo {author} {\bibfnamefont {G.}~\bibnamefont {Heinrich}},
  \bibinfo {author} {\bibfnamefont {T.}~\bibnamefont {Huber}}, \ and\ \bibinfo
  {author} {\bibfnamefont {C.}~\bibnamefont {Studerus}},\ }\href {\doibase
  10.1016/j.physletb.2006.08.008} {\bibfield  {journal} {\bibinfo  {journal}
  {Phys. Lett.}\ }\textbf {\bibinfo {volume} {B640}},\ \bibinfo {pages} {252}
  (\bibinfo {year} {2006})}\BibitemShut {NoStop}%
\bibitem [{\citenamefont {Heinrich}\ \emph {et~al.}(2008)\citenamefont
  {Heinrich}, \citenamefont {Huber},\ and\ \citenamefont
  {Maitre}}]{Heinrich:2007at}%
  \BibitemOpen
  \bibfield  {author} {\bibinfo {author} {\bibfnamefont {G.}~\bibnamefont
  {Heinrich}}, \bibinfo {author} {\bibfnamefont {T.}~\bibnamefont {Huber}}, \
  and\ \bibinfo {author} {\bibfnamefont {D.}~\bibnamefont {Maitre}},\ }\href
  {\doibase 10.1016/j.physletb.2008.03.028} {\bibfield  {journal} {\bibinfo
  {journal} {Phys. Lett.}\ }\textbf {\bibinfo {volume} {B662}},\ \bibinfo
  {pages} {344} (\bibinfo {year} {2008})}\BibitemShut {NoStop}%
\bibitem [{\citenamefont {Heinrich}\ \emph {et~al.}(2009)\citenamefont
  {Heinrich}, \citenamefont {Huber}, \citenamefont {Kosower},\ and\
  \citenamefont {Smirnov}}]{Heinrich:2009be}%
  \BibitemOpen
  \bibfield  {author} {\bibinfo {author} {\bibfnamefont {G.}~\bibnamefont
  {Heinrich}}, \bibinfo {author} {\bibfnamefont {T.}~\bibnamefont {Huber}},
  \bibinfo {author} {\bibfnamefont {D.~A.}\ \bibnamefont {Kosower}}, \ and\
  \bibinfo {author} {\bibfnamefont {V.~A.}\ \bibnamefont {Smirnov}},\ }\href
  {\doibase 10.1016/j.physletb.2009.06.038} {\bibfield  {journal} {\bibinfo
  {journal} {Phys. Lett.}\ }\textbf {\bibinfo {volume} {B678}},\ \bibinfo
  {pages} {359} (\bibinfo {year} {2009})}\BibitemShut {NoStop}%
\bibitem [{\citenamefont {Lee}\ \emph {et~al.}(2010{\natexlab{a}})\citenamefont
  {Lee}, \citenamefont {Smirnov},\ and\ \citenamefont {Smirnov}}]{Lee:2010cga}%
  \BibitemOpen
  \bibfield  {author} {\bibinfo {author} {\bibfnamefont {R.~N.}\ \bibnamefont
  {Lee}}, \bibinfo {author} {\bibfnamefont {A.~V.}\ \bibnamefont {Smirnov}}, \
  and\ \bibinfo {author} {\bibfnamefont {V.~A.}\ \bibnamefont {Smirnov}},\
  }\href {\doibase 10.1007/JHEP04(2010)020} {\bibfield  {journal} {\bibinfo
  {journal} {JHEP}\ }\textbf {\bibinfo {volume} {04}},\ \bibinfo {pages} {020}
  (\bibinfo {year} {2010}{\natexlab{a}})}\BibitemShut {NoStop}%
\bibitem [{\citenamefont {Lee}\ and\ \citenamefont
  {Smirnov}(2011)}]{Lee:2010ik}%
  \BibitemOpen
  \bibfield  {author} {\bibinfo {author} {\bibfnamefont {R.~N.}\ \bibnamefont
  {Lee}}\ and\ \bibinfo {author} {\bibfnamefont {V.~A.}\ \bibnamefont
  {Smirnov}},\ }\href {\doibase 10.1007/JHEP02(2011)102} {\bibfield  {journal}
  {\bibinfo  {journal} {JHEP}\ }\textbf {\bibinfo {volume} {02}},\ \bibinfo
  {pages} {102} (\bibinfo {year} {2011})}\BibitemShut {NoStop}%
\bibitem [{\citenamefont {Lee}\ \emph {et~al.}(2010{\natexlab{b}})\citenamefont
  {Lee}, \citenamefont {Smirnov},\ and\ \citenamefont {Smirnov}}]{Lee:2010ug}%
  \BibitemOpen
  \bibfield  {author} {\bibinfo {author} {\bibfnamefont {R.~N.}\ \bibnamefont
  {Lee}}, \bibinfo {author} {\bibfnamefont {A.~V.}\ \bibnamefont {Smirnov}}, \
  and\ \bibinfo {author} {\bibfnamefont {V.~A.}\ \bibnamefont {Smirnov}},\
  }\href {\doibase 10.1016/j.nuclphysbps.2010.09.011} {\bibfield  {journal}
  {\bibinfo  {journal} {Nucl. Phys. Proc. Suppl.}\ }\textbf {\bibinfo {volume}
  {205-206}},\ \bibinfo {pages} {308} (\bibinfo {year}
  {2010}{\natexlab{b}})}\BibitemShut {NoStop}%
\bibitem [{\citenamefont {Gehrmann}\ \emph {et~al.}(2010)\citenamefont
  {Gehrmann}, \citenamefont {Glover}, \citenamefont {Huber}, \citenamefont
  {Ikizlerli},\ and\ \citenamefont {Studerus}}]{Gehrmann:2010ue}%
  \BibitemOpen
  \bibfield  {author} {\bibinfo {author} {\bibfnamefont {T.}~\bibnamefont
  {Gehrmann}}, \bibinfo {author} {\bibfnamefont {E.~W.~N.}\ \bibnamefont
  {Glover}}, \bibinfo {author} {\bibfnamefont {T.}~\bibnamefont {Huber}},
  \bibinfo {author} {\bibfnamefont {N.}~\bibnamefont {Ikizlerli}}, \ and\
  \bibinfo {author} {\bibfnamefont {C.}~\bibnamefont {Studerus}},\ }\href
  {\doibase 10.1007/JHEP06(2010)094} {\bibfield  {journal} {\bibinfo  {journal}
  {JHEP}\ }\textbf {\bibinfo {volume} {06}},\ \bibinfo {pages} {094} (\bibinfo
  {year} {2010})}\BibitemShut {NoStop}%
\bibitem [{\citenamefont {Ravindran}(2006{\natexlab{a}})}]{Ravindran:2005vv}%
  \BibitemOpen
  \bibfield  {author} {\bibinfo {author} {\bibfnamefont {V.}~\bibnamefont
  {Ravindran}},\ }\href {\doibase 10.1016/j.nuclphysb.2006.04.008} {\bibfield
  {journal} {\bibinfo  {journal} {Nucl.Phys.}\ }\textbf {\bibinfo {volume}
  {B746}},\ \bibinfo {pages} {58} (\bibinfo {year}
  {2006}{\natexlab{a}})}\BibitemShut {NoStop}%
%%CITATION = HEP-PH/0512249;%%
\bibitem [{\citenamefont {Ravindran}(2006{\natexlab{b}})}]{Ravindran:2006cg}%
  \BibitemOpen
  \bibfield  {author} {\bibinfo {author} {\bibfnamefont {V.}~\bibnamefont
  {Ravindran}},\ }\href {\doibase 10.1016/j.nuclphysb.2006.06.025} {\bibfield
  {journal} {\bibinfo  {journal} {Nucl. Phys.}\ }\textbf {\bibinfo {volume}
  {B752}},\ \bibinfo {pages} {173} (\bibinfo {year}
  {2006}{\natexlab{b}})}\BibitemShut {NoStop}%
\bibitem [{\citenamefont {Korchemsky}\ and\ \citenamefont
  {Radyushkin}(1987)}]{Korchemsky:1987wg}%
  \BibitemOpen
  \bibfield  {author} {\bibinfo {author} {\bibfnamefont {G.~P.}\ \bibnamefont
  {Korchemsky}}\ and\ \bibinfo {author} {\bibfnamefont {A.~V.}\ \bibnamefont
  {Radyushkin}},\ }\href {\doibase 10.1016/0550-3213(87)90277-X} {\bibfield
  {journal} {\bibinfo  {journal} {Nucl. Phys.}\ }\textbf {\bibinfo {volume}
  {B283}},\ \bibinfo {pages} {342} (\bibinfo {year} {1987})}\BibitemShut
  {NoStop}%
%%CITATION = NUPHA,B283,342;%%
\bibitem [{\citenamefont {Correa}\ \emph {et~al.}(2012)\citenamefont {Correa},
  \citenamefont {Henn}, \citenamefont {Maldacena},\ and\ \citenamefont
  {Sever}}]{Correa:2012nk}%
  \BibitemOpen
  \bibfield  {author} {\bibinfo {author} {\bibfnamefont {D.}~\bibnamefont
  {Correa}}, \bibinfo {author} {\bibfnamefont {J.}~\bibnamefont {Henn}},
  \bibinfo {author} {\bibfnamefont {J.}~\bibnamefont {Maldacena}}, \ and\
  \bibinfo {author} {\bibfnamefont {A.}~\bibnamefont {Sever}},\ }\href
  {\doibase 10.1007/JHEP05(2012)098} {\bibfield  {journal} {\bibinfo  {journal}
  {JHEP}\ }\textbf {\bibinfo {volume} {05}},\ \bibinfo {pages} {098} (\bibinfo
  {year} {2012})}\BibitemShut {NoStop}%
%%CITATION = ARXIV:1203.1019;%%
\bibitem [{\citenamefont {Moch}\ \emph {et~al.}(2004)\citenamefont {Moch},
  \citenamefont {Vermaseren},\ and\ \citenamefont {Vogt}}]{Moch:2004pa}%
  \BibitemOpen
  \bibfield  {author} {\bibinfo {author} {\bibfnamefont {S.}~\bibnamefont
  {Moch}}, \bibinfo {author} {\bibfnamefont {J.}~\bibnamefont {Vermaseren}}, \
  and\ \bibinfo {author} {\bibfnamefont {A.}~\bibnamefont {Vogt}},\ }\href
  {\doibase 10.1016/j.nuclphysb.2004.03.030} {\bibfield  {journal} {\bibinfo
  {journal} {Nucl.Phys.}\ }\textbf {\bibinfo {volume} {B688}},\ \bibinfo
  {pages} {101} (\bibinfo {year} {2004})}\BibitemShut {NoStop}%
\bibitem [{\citenamefont {Vogt}\ \emph {et~al.}(2004)\citenamefont {Vogt},
  \citenamefont {Moch},\ and\ \citenamefont {Vermaseren}}]{Vogt:2004mw}%
  \BibitemOpen
  \bibfield  {author} {\bibinfo {author} {\bibfnamefont {A.}~\bibnamefont
  {Vogt}}, \bibinfo {author} {\bibfnamefont {S.}~\bibnamefont {Moch}}, \ and\
  \bibinfo {author} {\bibfnamefont {J.}~\bibnamefont {Vermaseren}},\ }\href
  {\doibase 10.1016/j.nuclphysb.2004.04.024} {\bibfield  {journal} {\bibinfo
  {journal} {Nucl.Phys.}\ }\textbf {\bibinfo {volume} {B691}},\ \bibinfo
  {pages} {129} (\bibinfo {year} {2004})}\BibitemShut {NoStop}%
\bibitem [{\citenamefont {Vogt}(2001)}]{Vogt:2000ci}%
  \BibitemOpen
  \bibfield  {author} {\bibinfo {author} {\bibfnamefont {A.}~\bibnamefont
  {Vogt}},\ }\href {\doibase 10.1016/S0370-2693(00)01344-7} {\bibfield
  {journal} {\bibinfo  {journal} {Phys.Lett.}\ }\textbf {\bibinfo {volume}
  {B497}},\ \bibinfo {pages} {228} (\bibinfo {year} {2001})}\BibitemShut
  {NoStop}%
\bibitem [{\citenamefont {Anselmi}\ \emph {et~al.}(1997)\citenamefont
  {Anselmi}, \citenamefont {Grisaru},\ and\ \citenamefont
  {Johansen}}]{Anselmi:1996mq}%
  \BibitemOpen
  \bibfield  {author} {\bibinfo {author} {\bibfnamefont {D.}~\bibnamefont
  {Anselmi}}, \bibinfo {author} {\bibfnamefont {M.~T.}\ \bibnamefont
  {Grisaru}}, \ and\ \bibinfo {author} {\bibfnamefont {A.}~\bibnamefont
  {Johansen}},\ }\href {\doibase 10.1016/S0550-3213(97)00108-9} {\bibfield
  {journal} {\bibinfo  {journal} {Nucl. Phys.}\ }\textbf {\bibinfo {volume}
  {B491}},\ \bibinfo {pages} {221} (\bibinfo {year} {1997})}\BibitemShut
  {NoStop}%
%%CITATION = HEP-TH/9601023;%%
\bibitem [{\citenamefont {Eden}\ \emph {et~al.}(2000)\citenamefont {Eden},
  \citenamefont {Schubert},\ and\ \citenamefont {Sokatchev}}]{Eden:2000mv}%
  \BibitemOpen
  \bibfield  {author} {\bibinfo {author} {\bibfnamefont {B.}~\bibnamefont
  {Eden}}, \bibinfo {author} {\bibfnamefont {C.}~\bibnamefont {Schubert}}, \
  and\ \bibinfo {author} {\bibfnamefont {E.}~\bibnamefont {Sokatchev}},\ }\href
  {\doibase 10.1016/S0370-2693(00)00515-3} {\bibfield  {journal} {\bibinfo
  {journal} {Phys. Lett.}\ }\textbf {\bibinfo {volume} {B482}},\ \bibinfo
  {pages} {309} (\bibinfo {year} {2000})}\BibitemShut {NoStop}%
%%CITATION = HEP-TH/0003096;%%
\bibitem [{\citenamefont {Bianchi}\ \emph {et~al.}(2000)\citenamefont
  {Bianchi}, \citenamefont {Kovacs}, \citenamefont {Rossi},\ and\ \citenamefont
  {Stanev}}]{Bianchi:2000hn}%
  \BibitemOpen
  \bibfield  {author} {\bibinfo {author} {\bibfnamefont {M.}~\bibnamefont
  {Bianchi}}, \bibinfo {author} {\bibfnamefont {S.}~\bibnamefont {Kovacs}},
  \bibinfo {author} {\bibfnamefont {G.}~\bibnamefont {Rossi}}, \ and\ \bibinfo
  {author} {\bibfnamefont {Y.~S.}\ \bibnamefont {Stanev}},\ }\href {\doibase
  10.1016/S0550-3213(00)00312-6} {\bibfield  {journal} {\bibinfo  {journal}
  {Nucl. Phys.}\ }\textbf {\bibinfo {volume} {B584}},\ \bibinfo {pages} {216}
  (\bibinfo {year} {2000})}\BibitemShut {NoStop}%
%%CITATION = HEP-TH/0003203;%%
\bibitem [{\citenamefont {Kotikov}\ \emph {et~al.}(2004)\citenamefont
  {Kotikov}, \citenamefont {Lipatov}, \citenamefont {Onishchenko},\ and\
  \citenamefont {Velizhanin}}]{Kotikov:2004er}%
  \BibitemOpen
  \bibfield  {author} {\bibinfo {author} {\bibfnamefont {A.~V.}\ \bibnamefont
  {Kotikov}}, \bibinfo {author} {\bibfnamefont {L.~N.}\ \bibnamefont
  {Lipatov}}, \bibinfo {author} {\bibfnamefont {A.~I.}\ \bibnamefont
  {Onishchenko}}, \ and\ \bibinfo {author} {\bibfnamefont {V.~N.}\ \bibnamefont
  {Velizhanin}},\ }\href {\doibase 10.1016/j.physletb.2004.05.078,
  10.1016/j.physletb.2005.11.002} {\bibfield  {journal} {\bibinfo  {journal}
  {Phys. Lett.}\ }\textbf {\bibinfo {volume} {B595}},\ \bibinfo {pages} {521}
  (\bibinfo {year} {2004})},\ \bibinfo {note} {[Erratum: Phys.
  Lett.B632,754(2006)]}\BibitemShut {NoStop}%
%%CITATION = HEP-TH/0404092;%%
\bibitem [{\citenamefont {Eden}\ \emph {et~al.}(2005)\citenamefont {Eden},
  \citenamefont {Jarczak},\ and\ \citenamefont {Sokatchev}}]{Eden:2004ua}%
  \BibitemOpen
  \bibfield  {author} {\bibinfo {author} {\bibfnamefont {B.}~\bibnamefont
  {Eden}}, \bibinfo {author} {\bibfnamefont {C.}~\bibnamefont {Jarczak}}, \
  and\ \bibinfo {author} {\bibfnamefont {E.}~\bibnamefont {Sokatchev}},\ }\href
  {\doibase 10.1016/j.nuclphysb.2005.01.036} {\bibfield  {journal} {\bibinfo
  {journal} {Nucl. Phys.}\ }\textbf {\bibinfo {volume} {B712}},\ \bibinfo
  {pages} {157} (\bibinfo {year} {2005})}\BibitemShut {NoStop}%
%%CITATION = HEP-TH/0409009;%%
\bibitem [{\citenamefont {Bern}\ \emph {et~al.}(2007)\citenamefont {Bern},
  \citenamefont {Czakon}, \citenamefont {Dixon}, \citenamefont {Kosower},\ and\
  \citenamefont {Smirnov}}]{Bern:2006ew}%
  \BibitemOpen
  \bibfield  {author} {\bibinfo {author} {\bibfnamefont {Z.}~\bibnamefont
  {Bern}}, \bibinfo {author} {\bibfnamefont {M.}~\bibnamefont {Czakon}},
  \bibinfo {author} {\bibfnamefont {L.~J.}\ \bibnamefont {Dixon}}, \bibinfo
  {author} {\bibfnamefont {D.~A.}\ \bibnamefont {Kosower}}, \ and\ \bibinfo
  {author} {\bibfnamefont {V.~A.}\ \bibnamefont {Smirnov}},\ }\href {\doibase
  10.1103/PhysRevD.75.085010} {\bibfield  {journal} {\bibinfo  {journal} {Phys.
  Rev.}\ }\textbf {\bibinfo {volume} {D75}},\ \bibinfo {pages} {085010}
  (\bibinfo {year} {2007})}\BibitemShut {NoStop}%
%%CITATION = HEP-TH/0610248;%%
\bibitem [{\citenamefont {Cachazo}\ \emph
  {et~al.}(2007{\natexlab{a}})\citenamefont {Cachazo}, \citenamefont
  {Spradlin},\ and\ \citenamefont {Volovich}}]{Cachazo:2006az}%
  \BibitemOpen
  \bibfield  {author} {\bibinfo {author} {\bibfnamefont {F.}~\bibnamefont
  {Cachazo}}, \bibinfo {author} {\bibfnamefont {M.}~\bibnamefont {Spradlin}}, \
  and\ \bibinfo {author} {\bibfnamefont {A.}~\bibnamefont {Volovich}},\ }\href
  {\doibase 10.1103/PhysRevD.75.105011} {\bibfield  {journal} {\bibinfo
  {journal} {Phys. Rev.}\ }\textbf {\bibinfo {volume} {D75}},\ \bibinfo {pages}
  {105011} (\bibinfo {year} {2007}{\natexlab{a}})}\BibitemShut {NoStop}%
%%CITATION = HEP-TH/0612309;%%
\bibitem [{\citenamefont {Henn}\ \emph {et~al.}(2010)\citenamefont {Henn},
  \citenamefont {Naculich}, \citenamefont {Schnitzer},\ and\ \citenamefont
  {Spradlin}}]{Henn:2010ir}%
  \BibitemOpen
  \bibfield  {author} {\bibinfo {author} {\bibfnamefont {J.~M.}\ \bibnamefont
  {Henn}}, \bibinfo {author} {\bibfnamefont {S.~G.}\ \bibnamefont {Naculich}},
  \bibinfo {author} {\bibfnamefont {H.~J.}\ \bibnamefont {Schnitzer}}, \ and\
  \bibinfo {author} {\bibfnamefont {M.}~\bibnamefont {Spradlin}},\ }\href
  {\doibase 10.1007/JHEP08(2010)002} {\bibfield  {journal} {\bibinfo  {journal}
  {JHEP}\ }\textbf {\bibinfo {volume} {08}},\ \bibinfo {pages} {002} (\bibinfo
  {year} {2010})}\BibitemShut {NoStop}%
%%CITATION = ARXIV:1004.5381;%%
\bibitem [{\citenamefont {Henn}\ and\ \citenamefont
  {Huber}(2013)}]{Henn:2013wfa}%
  \BibitemOpen
  \bibfield  {author} {\bibinfo {author} {\bibfnamefont {J.~M.}\ \bibnamefont
  {Henn}}\ and\ \bibinfo {author} {\bibfnamefont {T.}~\bibnamefont {Huber}},\
  }\href {\doibase 10.1007/JHEP09(2013)147} {\bibfield  {journal} {\bibinfo
  {journal} {JHEP}\ }\textbf {\bibinfo {volume} {09}},\ \bibinfo {pages} {147}
  (\bibinfo {year} {2013})}\BibitemShut {NoStop}%
%%CITATION = ARXIV:1304.6418;%%
\bibitem [{\citenamefont {Beisert}\ \emph {et~al.}(2007)\citenamefont
  {Beisert}, \citenamefont {Eden},\ and\ \citenamefont
  {Staudacher}}]{Beisert:2006ez}%
  \BibitemOpen
  \bibfield  {author} {\bibinfo {author} {\bibfnamefont {N.}~\bibnamefont
  {Beisert}}, \bibinfo {author} {\bibfnamefont {B.}~\bibnamefont {Eden}}, \
  and\ \bibinfo {author} {\bibfnamefont {M.}~\bibnamefont {Staudacher}},\
  }\href {\doibase 10.1088/1742-5468/2007/01/P01021} {\bibfield  {journal}
  {\bibinfo  {journal} {J. Stat. Mech.}\ }\textbf {\bibinfo {volume} {0701}},\
  \bibinfo {pages} {P01021} (\bibinfo {year} {2007})}\BibitemShut {NoStop}%
%%CITATION = HEP-TH/0610251;%%
\bibitem [{\citenamefont {Cachazo}\ \emph
  {et~al.}(2007{\natexlab{b}})\citenamefont {Cachazo}, \citenamefont
  {Spradlin},\ and\ \citenamefont {Volovich}}]{Cachazo:2007ad}%
  \BibitemOpen
  \bibfield  {author} {\bibinfo {author} {\bibfnamefont {F.}~\bibnamefont
  {Cachazo}}, \bibinfo {author} {\bibfnamefont {M.}~\bibnamefont {Spradlin}}, \
  and\ \bibinfo {author} {\bibfnamefont {A.}~\bibnamefont {Volovich}},\ }\href
  {\doibase 10.1103/PhysRevD.76.106004} {\bibfield  {journal} {\bibinfo
  {journal} {Phys. Rev.}\ }\textbf {\bibinfo {volume} {D76}},\ \bibinfo {pages}
  {106004} (\bibinfo {year} {2007}{\natexlab{b}})}\BibitemShut {NoStop}%
%%CITATION = ARXIV:0707.1903;%%
\bibitem [{\citenamefont {Bern}\ \emph {et~al.}(2008)\citenamefont {Bern},
  \citenamefont {Dixon}, \citenamefont {Kosower}, \citenamefont {Roiban},
  \citenamefont {Spradlin}, \citenamefont {Vergu},\ and\ \citenamefont
  {Volovich}}]{Bern:2008ap}%
  \BibitemOpen
  \bibfield  {author} {\bibinfo {author} {\bibfnamefont {Z.}~\bibnamefont
  {Bern}}, \bibinfo {author} {\bibfnamefont {L.~J.}\ \bibnamefont {Dixon}},
  \bibinfo {author} {\bibfnamefont {D.~A.}\ \bibnamefont {Kosower}}, \bibinfo
  {author} {\bibfnamefont {R.}~\bibnamefont {Roiban}}, \bibinfo {author}
  {\bibfnamefont {M.}~\bibnamefont {Spradlin}}, \bibinfo {author}
  {\bibfnamefont {C.}~\bibnamefont {Vergu}}, \ and\ \bibinfo {author}
  {\bibfnamefont {A.}~\bibnamefont {Volovich}},\ }\href {\doibase
  10.1103/PhysRevD.78.045007} {\bibfield  {journal} {\bibinfo  {journal} {Phys.
  Rev.}\ }\textbf {\bibinfo {volume} {D78}},\ \bibinfo {pages} {045007}
  (\bibinfo {year} {2008})}\BibitemShut {NoStop}%
\bibitem [{\citenamefont {Fiamberti}\ \emph
  {et~al.}(2008{\natexlab{a}})\citenamefont {Fiamberti}, \citenamefont
  {Santambrogio}, \citenamefont {Sieg},\ and\ \citenamefont
  {Zanon}}]{Fiamberti:2007rj}%
  \BibitemOpen
  \bibfield  {author} {\bibinfo {author} {\bibfnamefont {F.}~\bibnamefont
  {Fiamberti}}, \bibinfo {author} {\bibfnamefont {A.}~\bibnamefont
  {Santambrogio}}, \bibinfo {author} {\bibfnamefont {C.}~\bibnamefont {Sieg}},
  \ and\ \bibinfo {author} {\bibfnamefont {D.}~\bibnamefont {Zanon}},\ }\href
  {\doibase 10.1016/j.physletb.2008.06.061} {\bibfield  {journal} {\bibinfo
  {journal} {Phys. Lett.}\ }\textbf {\bibinfo {volume} {B666}},\ \bibinfo
  {pages} {100} (\bibinfo {year} {2008}{\natexlab{a}})}\BibitemShut {NoStop}%
%%CITATION = ARXIV:0712.3522;%%
\bibitem [{\citenamefont {Fiamberti}\ \emph
  {et~al.}(2008{\natexlab{b}})\citenamefont {Fiamberti}, \citenamefont
  {Santambrogio}, \citenamefont {Sieg},\ and\ \citenamefont
  {Zanon}}]{Fiamberti:2008sh}%
  \BibitemOpen
  \bibfield  {author} {\bibinfo {author} {\bibfnamefont {F.}~\bibnamefont
  {Fiamberti}}, \bibinfo {author} {\bibfnamefont {A.}~\bibnamefont
  {Santambrogio}}, \bibinfo {author} {\bibfnamefont {C.}~\bibnamefont {Sieg}},
  \ and\ \bibinfo {author} {\bibfnamefont {D.}~\bibnamefont {Zanon}},\ }\href
  {\doibase 10.1016/j.nuclphysb.2008.07.014} {\bibfield  {journal} {\bibinfo
  {journal} {Nucl. Phys.}\ }\textbf {\bibinfo {volume} {B805}},\ \bibinfo
  {pages} {231} (\bibinfo {year} {2008}{\natexlab{b}})}\BibitemShut {NoStop}%
%%CITATION = ARXIV:0806.2095;%%
\bibitem [{\citenamefont {Velizhanin}(2009{\natexlab{a}})}]{Velizhanin:2008jd}%
  \BibitemOpen
  \bibfield  {author} {\bibinfo {author} {\bibfnamefont {V.~N.}\ \bibnamefont
  {Velizhanin}},\ }\href {\doibase 10.1134/S0021364009010020} {\bibfield
  {journal} {\bibinfo  {journal} {JETP Lett.}\ }\textbf {\bibinfo {volume}
  {89}},\ \bibinfo {pages} {6} (\bibinfo {year}
  {2009}{\natexlab{a}})}\BibitemShut {NoStop}%
%%CITATION = ARXIV:0808.3832;%%
\bibitem [{\citenamefont {Velizhanin}(2009{\natexlab{b}})}]{Velizhanin:2009gv}%
  \BibitemOpen
  \bibfield  {author} {\bibinfo {author} {\bibfnamefont {V.~N.}\ \bibnamefont
  {Velizhanin}},\ }\href {\doibase 10.1134/S0021364009120017} {\bibfield
  {journal} {\bibinfo  {journal} {JETP Lett.}\ }\textbf {\bibinfo {volume}
  {89}},\ \bibinfo {pages} {593} (\bibinfo {year}
  {2009}{\natexlab{b}})}\BibitemShut {NoStop}%
%%CITATION = ARXIV:0902.4646;%%
\end{thebibliography}%
\bibliographystyle{apsrev4-1}
\end{document}